\documentclass[lettersize,journal]{IEEEtran}
\usepackage{amsmath,amsfonts,amssymb}
\usepackage{booktabs}
\usepackage{algorithm}
\usepackage{algorithmic}
\usepackage{array}
\usepackage[caption=false,font=normalsize,labelfont=sf,textfont=sf]{subfig}
\usepackage[justification=raggedright]{caption}
\usepackage{textcomp}
\usepackage{stfloats}
\usepackage{url}
\usepackage{verbatim}
\usepackage{graphicx}
\usepackage{subfig}
\usepackage[style=ieee,backend=bibtex]{biblatex}
\usepackage{wrapfig}
\usepackage{tabularx}
\usepackage{enumitem,kantlipsum}
\DeclareFieldFormat{doi}{} 
\usepackage{cleveref}
\crefname{figure}{Fig.}{Figs.}
\Crefname{figure}{Fig.}{Figs.}
\crefname{equation}{Eq.}{Eqs.}
\Crefname{equation}{Eq.}{Eqs.}
\usepackage{xcolor}
\usepackage{soul}
\soulregister\cite7 
\soulregister\ref7
\soulregister\eqref7

\begin{document}
\captionsetup[subfigure]{font=small, labelfont=small} 
\title{Precise Near-Field Beam Training with DFT Codebook based on Amplitude-only Measurement}
\author{Zijun Wang,  Shawn Tsai, Rama Kiran, Rui Zhang,~\IEEEmembership{member,~IEEE}
\thanks{This paper was produced by the IEEE Publication Technology Group. They are in Piscataway, NJ.}
\thanks{Zijun Wang, Rui Zhang are with the Department of Electrical Engineering, The State University of New York at Buffalo, New York, USA  (email: {zwang267@buffalo.edu}, {rzhang45@buffalo.edu}).}
\thanks{Shawn Tsai is with the CSD, MediaTek Inc. USA, San Diego, CA 92122 USA (e-mail: shawn.tsai@mediatek.com). }
\thanks{Rama Kiran is with CSD, MediaTek Inc., Bengaluru, India. (e-mail: rama.kiran@mediatek.com) }
\thanks{Corresponding author: Rui Zhang.}.
}

\markboth{Journal of \LaTeX\ Class Files,~Vol.~14, No.~8, August~2021}%
{Shell \MakeLowercase{\textit{et al.}}: A Sample Article Using IEEEtran.cls for IEEE Journals}

\IEEEpubid{0000--0000/00\$00.00~\copyright~2025 IEEE}

\maketitle
\begin{abstract}
Extremely large antenna arrays (ELAAs) operating in high-frequency bands have spurred the development of near-field communication, driving advancements in beam training and signal processing design.  In this work, we present a low-complexity near-field beam training scheme that fully utilize the conventional discrete Fourier transform (DFT) codebook designed for far-field users. We begin by analyzing the received beam pattern in the near field and derive closed-form expressions for the beam width and central gain. These analytical results enable the definition of an angle-dependent, modified Rayleigh distance, which effectively distinguishes near-field and far-field user regimes. Building on the analysis, we develop a direct and computationally efficient method to estimate user distance, with a complexity of $\mathcal{O}(1)$, and further improve its accuracy through a simple refinement. Simulation results demonstrate significant gains in both single- and multi-user settings, with up to 2.38 dB SNR improvement over exhaustive search. To further enhance estimation accuracy, we additionally propose a maximum likelihood estimation (MLE) based refinement method, leveraging the Rician distribution of signal amplitudes and achieving accuracy close to the Cramér–Rao bound (CRB). Simulation shows the single-user and multi-user achievable rates can both approach those obtained with ideal channel state information.
\end{abstract}


\section{Introduction}
\IEEEPARstart{T}{he}-{imminent advent of the 6G era envisions a wireless network utilizing high-frequency (RF) bands, such as millimeter waves and terahertz frequencies, to address the escalating demand for system capacity driven by emerging intelligent applications and services (e.g., AR, VR, etc)\cite{6G-tutorial,6G-tutorial2}. {To effectively compensate for high‐frequency propagation losses, extremely large antenna arrays (ELAA) comprising hundreds or even thousands of elements are expected to be deployed }\cite{terahertz-tutorial,mimo-tutorial}. 
However, the combination of large antenna arrays and shorter wavelengths {introduces near-field} effect \cite{Tutorial_with_beam_pattern_and_resolution, General_tutorial}. The near-field regime arises when the communication distance is on the order of, or smaller than, the Rayleigh distance threshold \cite{fra_and_ray_2}. Traditional wireless communication system design is based on far-field assumptions, wherein electromagnetic (EM) waves are approximated as planar \cite{far-field_assumption}. In this region, the signal at each antenna element differs primarily by a phase shift determined solely by the angle of arrival \cite{far-field-assumption2}. Conversely, in the near-field region the EM wavefront is spherical, the steering vector—previously determined solely by angle—now includes both angular and distance parameters for near field users \cite{Cui_and_Dai_channel_model, Fraunhofer_and_Fresnel_Distances,channel_model2}.
\IEEEpubidadjcol 

This fundamental shift demands new beamforming and beam‐training strategies designed specifically for near‐field communications. Recent research has introduced dedicated channel estimation, beam training, and beamforming techniques tailored to the near‐field regime.  \cite{Beam_focusing_multiuser} {optimize multi-user MIMO beam forming considering different antenna structures with the assumption that the precise user locations are known.}
A polar-domain codebook that integrates both angular and distance information is also developed, which is used in uplink channel estimation \cite{Cui_and_Dai_channel_model}. {However, exhaustive beam sweeping using the near-field codebook in \cite{Cui_and_Dai_channel_model} incurs a relatively higher complexity due to its enlarged codebook size. Consequently, there have been works focused on reducing either the near-field codebook size or the beam-sweeping overhead \cite{Hierarchical_near-field_beam_relocate,beam_split_wideband_beam_training, sequence_low_overhead_orth,k-b_domain_codebook}. } For example, \cite{Hierarchical_near-field_beam_relocate} proposes a two-layer polar-domain codebook, where the upper layer reduces training overhead by introducing angle- and distance-dependent phase factor to far-field codewords for coarse spatial search. Apart from designing hierarchical codebooks to reduce beam-sweeping overhead, \cite{beam_split_wideband_beam_training, sequence_low_overhead_orth} reduce codeword correlation in the polar domain via structural modifications such as frequency-dependent beam split and orthogonality-enhancing sequence design, enabling simultaneous searching in multi-angles and -distances with reduced beam-sweeping overhead.
\cite{k-b_domain_codebook} proposes a slope-intercept domain (k–b) representation of near-field channels, where codewords are designed by decoupling spatial chirp parameters to reduce correlation and facilitate low-overhead hierarchical beam training. 

{Multi-stage frameworks that combine DFT and near-field codebooks offer another path to overhead reduction.} \cite{Fast_Near-Field_Beam_Training} uses the DFT codebook in the first stage to estimate the user’s angular direction, and then estimates distance in the second stage using a near-field codebook. Based on this, to further improve angular resolution or reduce training overhead, some works additionally exploit subarray-based processing \cite{sparse_DFT_subarray,Two-Stage_Hierarchical_Beam_Training,Efficient_Hybrid_Near-_and_Far-Field_Beam_Training,RIS_two-stage_subarray}. 
\cite{Two-Stage_Hierarchical_Beam_Training,Efficient_Hybrid_Near-_and_Far-Field_Beam_Training} divide the array into a central subarray to perform coarse angle estimation using a DFT codebook, then activate the full array to refine both angle and distance using a near-field codebook. 
\cite{sparse_DFT_subarray} adopts a similar process, but it inherits a sparse DFT codebook constructed by analyzing the beam pattern.

Furthermore, several studies have leveraged only the far-field DFT codebook for near-field beam training or user localization. In \cite{RIS_two-stage_subarray}, the training phase sequentially activates subarrays of a reconfigurable intelligent surface (RIS) and applies DFT codebooks to each subarray—exploiting local far-field conditions to estimate both angle and distance with minimal overhead. Building on this subarray-based low-overhead approach, \cite{Scalable_Near-Field_Localization_UPA} enhances near-field localization by exploiting angle drifts across partitioned arrays, enabling scalable positioning without requiring full-array near-field modeling. \cite{Joint_Angle_and_Range_Estimation} numerically analyzes beamwidth and beam gain to jointly estimate the angle and distance of near-field users, thereby reducing computational complexity.

{Most existing beam‐training schemes employ discretely sampled codebooks without explicitly modeling the structure of measurement noise. As a result, they exhibit a significant performance gap compared to systems with full channel state information, particularly under noise interference and limited resolution. High-accuracy beam training methods have also been developed to address noise and resolution limitations. Some approaches based on deep learning has been developed to enhance robutsness to noise \cite{DL_for_beam_training,low-overhead_DL, DL_OMP_hybrid}.} 
In parallel, several studies employ maximum likelihood estimation (MLE) for additional refinement. For example, \cite{Cui_and_Dai_channel_model} {apply Gaussian‐MLE to refine ULA channel estimates derived from a near‐field codebook under the assumption of full access to the received signal amplitude and phase.} Similarly, \cite{Scalable_Near-Field_Localization_UPA} extends these assumptions to MLE‐based refinement of user positioning in uniform planar array (UPA) systems.

While recent low-overhead beam-training schemes based on DFT codebooks achieve reduced complexity, they often depend on simulation-driven heuristics. An analytical characterization of near-field beam patterns using DFT codebook would better close the gap in fundamental understanding of their structural properties for more efficient training. Moreover, existing MLE-based refinements assume full access to the received signal’s phase—a requirement that is often impractical due to phase distortions from carrier-frequency offsets \cite{Downlink_phase_noise}.
{In this work, we overcome these limitations by developing a near-field beam-training scheme that (i) leverages closed-form beam-pattern expressions and (ii) relies solely on amplitude-only measurements for distance and angle estimation.} This paper is extended from a conference version\cite{wang2025lowcomplexitynearfieldbeamtraining}. {Whereas \cite{wang2025lowcomplexitynearfieldbeamtraining} derived a closed-form beamwidth for a fixed 6 dB threshold and proposed a non-iterative coarse estimation method, this paper generalizes the beamwidth expression to arbitrary thresholds and introduces a modified Rayleigh distance to render the algorithm adaptive to both near-field and far-field users. We further enhance accuracy through an iterative refinement procedure and conclude with an MLE-based refinement scheme that leverages the amplitude distribution of the received signal. These extensions collectively enable more flexible, robust, and precise near-field beam training.} The main contributions of this paper are summarized below:

1. We derive close-form expressions for the beam width and central gain of {beam pattern using} DFT codewords in near-field conditions, capturing their dependence on user angle and distance. Based on this, we introduce a modified Rayleigh distance that more accurately distinguishes far- and near-field regimes.

2. {Using the derived formulas, we propose a two-stage beam training method with low computational complexity}: first estimating angle via clustered DFT sweeping and median-based selection, then estimating distance through the observed beamwidth. An iterative refinement procedure is applied thereafter to further enhance accuracy. 

3. {To avoid phase-distortion effects, we formulate a maximum-likelihood estimator based solely on signal amplitudes under a joint Rician model. We solve the resulting nonconvex problem using a hybrid Simulated Annealing (SA)–adaptive moment estimation (Adam) optimizer, achieving performance close to the Cramér–Rao bound (CRB).}


The rest of the paper is organized as follows. In \Cref{sec:system model}, we present the system model. \Cref{sec:beam pattern} analyzes the received beam pattern and derives closed‐form expressions for both beamwidth and central gain. Building on this theoretical foundation, \Cref{sec:coarse and refine} proposes a low‐complexity two‐stage beam‐training method along with an iterative refinement procedure, and compares its performance against benchmark schemes through simulation. In \Cref{sec:Proposed Maximum Likelihood Scheme}, we introduce the amplitude‐only MLE beam‐training scheme, derive its corresponding Cramér–Rao bound, and evaluate its accuracy. Finally, \Cref{sec:clu} concludes the paper.

The used notations are listed here. Lower-case boldface letters represent vectors and upper-case boldface letters represent matrices or vector set, respectively. $\mathcal{N}\triangleq \{0,1,\dots, N-1\}$ is the integer set where $N$ is a positive integer. $\jmath$ is $\sqrt{-1}$. $\mathcal{CN}(0,\sigma^2)$ is the complex Gaussian distribution with 0 mean and $\sigma^2$ variance. $\lvert\cdot\lvert$ denotes the absolute operator or the cardinality of a set. The error function is defined as $\operatorname{erf}(x)=\frac{2}{\sqrt{\pi}} \int_{0}^{x} e^{-y^{2}} \operatorname{d} y$. $\operatorname{Arg}[\cdot]$ denotes the argument of a complex number. $I_0(\cdot)$ and $I_1(\cdot)$ are the modified Bessel functions of the first kind, of order zero and one, respectively. $\operatorname{max}(\cdot)$ chooses the maximum value of a set, while $\min(\cdot)$ chooses the minimum value.  $\operatorname{Range}(\cdot)=\operatorname{max}(\cdot)-\operatorname{min}(\cdot)$ is the range of set $\cdot$. $\partial$ to denote the partial derivative operator. $\nabla$ is the gradient operator. $\operatorname{Cov}(\cdot)$ is the covariance matrix of vector $\cdot$. $\mathbb{E}[\cdot]$ denotes the expectation.

\section{System Model}\label{sec:system model}
In this article, we consider a base station (BS) with the ULA consisting of $N$ antenna elements, and without loss of generality, user is equipped with one single antenna. The BS adopts DFT codebook to perform beam training.   For uniform linear arrays (ULA), conventional method defines the near-field region as between Fresnel distance $R_{Fre}=\frac{1}{2} \sqrt{\frac{D^{3}}{\lambda}}$ and Rayleigh distance $R_{Ray}=\frac{2D^2}{\lambda}$, in which $\lambda$ is the wavelength at the central frequency. $D$ is the aperture of ULA at BS \cite{Fraunhofer_and_Fresnel_Distances}. The far field lies farther away than $R_{Ray}$. 
\subsection{Channel Model}
We assume that the antenna is placed on the y-axis, with the center at $(0, 0)$, and each antenna element has the coordinate of $(0,\delta_{n} d)$, with $\delta_{n}=\frac{2n-N+1}{2},n\in\mathcal{N}$ being the index of the antenna elements. The diagram of antenna and user can be found in \Cref{fig:antenna}.
\begin{figure}[!t]
\captionsetup{justification=raggedright, singlelinecheck=false} 
\centering
\includegraphics[width=2.5in]{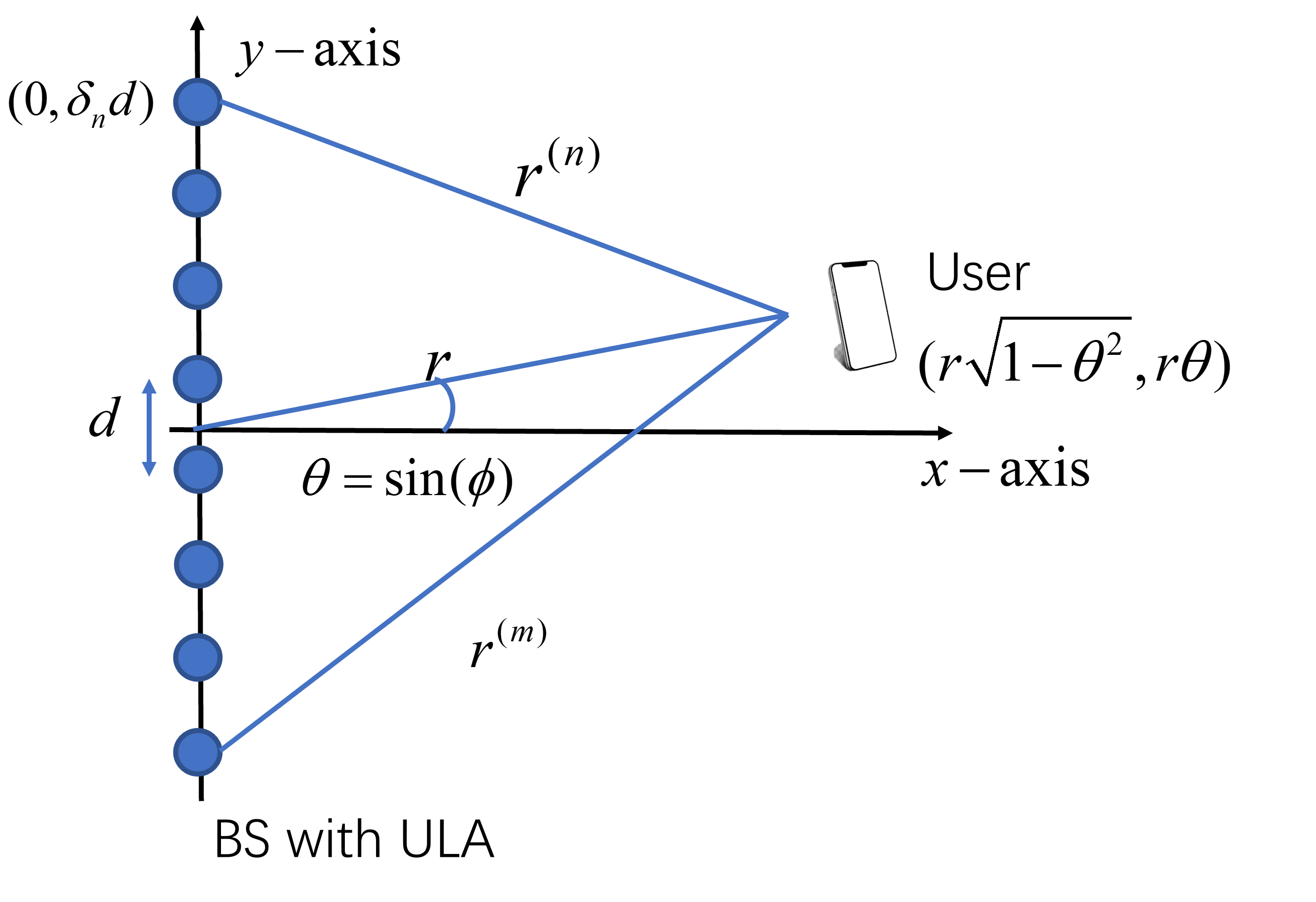}
\caption{Near-field communication system with ULA}
\label{fig:antenna}
\end{figure}
{Since high RF bands are dominated by line-of-sight (LoS) path\cite{The_Los_dominated_Tutorial}, We consider the following channel model
\begin{equation}
   \mathbf{h}^{H}\left(\theta, r\right)=\sqrt{N} g e^{-\frac{\jmath 2 \pi r}{\lambda}}\mathbf{b}^{H}(\theta, r),  
   \label{eq:channel model}
\end{equation}
where $g=\frac{\lambda}{4\pi r}$ is the channel gain. It represents the { path loss caused by distance} between BS and NUs or FUs, and
\begin{equation}
     \mathbf{b}(\theta, r)=\frac{1}{\sqrt{N}}\left[e^{-\jmath 2 \pi\left(r^{(0)}-r\right) / \lambda}, \cdots, e^{-\jmath 2 \pi\left(r^{(N-1)}-r\right) / \lambda}\right]^{T},
\end{equation}
is the near-field channel steering vector of ULA, in which $\theta=\sin (\phi)$ is the spatial user angle at BS, with range $[-1,1]$. $\phi$ is the angle-of-departure (AoD) of users. $r^{(n)}=\sqrt{r^{2}+\delta_{n}^{2} d^{2}-2 r \theta \delta_{n} d}$ is the physical distance from $n$-th antenna elements to users and {$r$ is the physical distance from central of the antenna array to users}.

\subsection{Received Signal Model}\label{subsec:signal}
For DFT codebook $\mathbf{V}_{DFT}=\{\mathbf{a}\left(\varphi_0 \right),\mathbf{a}\left(\varphi_1 \right),\cdots,\mathbf{a}\left(\varphi_{N-1} \right)\}$, we uniformly sample the angle domain such that $\varphi_{n}=\frac{2 n-N+1}{N},n\in \mathcal{N}$. Each codeword is denoted as \( \mathbf{v}_n = \mathbf{a}(\varphi_n) \).  The far-field codeword \( \mathbf{a}(\varphi_n) \) is defined as follows:
\begin{equation}
\mathbf{a}(\varphi_n) \triangleq \frac{1}{\sqrt{N}} 
\left[ 
e^{-j\pi \left(-\frac{N-1}{2}\right)\varphi_n},
\cdots,e^{-j\pi \left(\frac{N-1}{2}\right)\varphi_n}
\right],\forall n \in \mathcal{N}.
\end{equation}
Based on the channel model, we denote the received signal at a user with a position $(\theta,r)$ using beam forming vector $\mathbf{v}_n$ as
\begin{equation}
  y(\mathbf{v}_n)=\mathbf{h}^{H}\left(\theta, r\right)\mathbf{v}_n x+w,  
  \label{eq:received signal model}
\end{equation}
in which  
$w\sim\mathcal{CN}(0,\sigma^2)$ is the complex additive white Gaussian noise (AWGN). $x$ is the reference pilot signal, and we assume that $x=1$. 
\section{Received Beam Pattern Analysis}\label{sec:beam pattern}
This section analyzes the received beam pattern and derives the beam pattern width and gain as a function of distance and angle. Given user location \((\theta,r)\), we consider the following beam gain using a DFT codeword on NUs:
\begin{align}
    G\left(\mathbf{b}^{H}(\theta, r),  \mathbf{a}(\varphi) \right) &=\left|\mathbf{b}^{H}(\theta, r) \mathbf{a}(\varphi)\right|\nonumber\\
    & \approx\lvert f(\theta, r, \varphi)\rvert.
    \label{eq:G and f}
\end{align}
We use the Taylor series to approximate $r^{(n)}\approx r-\delta_{n} d \theta+\frac{\delta_{n}^{2} d^{2}\left(1-\theta^{2}\right)}{2 r}$. And we denote
\begin{align}
    &f(\theta, r, \varphi)\nonumber\\
    &=\frac{1}{N}\sum_{n=0}^{N-1} \exp \left(\jmath \pi\left[-\delta_{n}(\theta-\varphi)+\frac{\delta_{n}^{2} d\left(1-\theta^{2}\right)}{2 r}\right]\right),  
    \label{eq:f define}
\end{align}
where $\varphi \in \{\varphi_{n}\mid\varphi_{n}=\frac{2 n-N+1}{N},n\in \mathcal{N}\}$ is the angle sampled for DFT codeword and the NU location in the polar domain is denoted as \( (\theta, r) \in \mathbb{R}^2 \).
Let integral variable $x=\frac{2n-N+1}{N}\in[-1,1], n\in\mathcal{N}$, and the $f(\theta, r, \varphi)$ in \Cref{eq:f define} can be further approximated by converting the discrete sum into an integral as:
\begin{align} \label{eq:f_function2}
&\tilde{f}(\theta, r, \varphi) \nonumber\\
&= \frac{1}{2} \int_{-1}^{1} {\exp \left(\jmath \pi\left[\frac{N^{2} x^{2} d\left(1-\theta^{2}\right)}{8 r}-\frac{N x}{2}(\theta-\varphi)\right]\right) \operatorname{d} x}.  
\end{align}
The approximation in \Cref{eq:f_function2} requires $(\theta-\varphi)\ll2,\frac{Nd}{r}\ll2$ according to the Riemann Integral.
Then, we denote near-field focusing gain $\alpha=\frac{N^{2} d\left(1-\theta^{2}\right)}{8 r}$ and angular offset $\beta=\frac{N(\theta-\varphi)}{2}$, and plug them into \Cref{eq:f_function2}. The near-field focusing gain captures the essence of how the beam's focus sharpness is affected by the array configuration and user positioning and angular offset quantifies the angular deviation between the beam's steering direction and the user's actual direction with the specific antenna. Then, the function \( \tilde{f}(\theta, r, \varphi) \) becomes:
{\small
\begin{align}
    & \tilde{f}(\theta, r, \varphi)  \nonumber\\
   &=\frac{1}{2} \frac{e^{\jmath\left(\frac{-\beta^{2}+\alpha\pi}{4 \alpha}\right)}\left[\operatorname{erf}\left(\frac{e^{\jmath\frac{3}{4} \pi}\sqrt{\pi}(\beta-2 \alpha) }{2 \sqrt{\alpha}}\right)-\operatorname{erf}\left(\frac{e^{\jmath \frac{3}{4} \pi}\sqrt{\pi}(\beta+2 \alpha) }{2 \sqrt{\alpha}}\right)\right]}{2 \sqrt{\alpha}}.
    \label{eq:origin f}
\end{align}}
The detailed calculation can be found in \Cref{app:close-form}. In the following, we first derive the central beam pattern gain, and then use it to derive the beam width.
\subsection{Derivation of Central Beam Pattern Gain}\label{subsec:Central Beam Pattern Gain}
We define the central beam pattern gain as the value of $G\left(\mathbf{b}^{H}(\theta, r), \mathbf{a}(\varphi)\right)$ when $\varphi=\theta  $ (i.e., $\beta=0$). The \Cref{eq:origin f} can be re-written to 
\begin{equation}
    \tilde{f}(\theta, r, \theta)= - \frac{1}{2}\frac{e^{\jmath\frac{\pi}{4}} \, \mathrm{erf}\left[e^{\jmath\frac{3}{4}\pi} \sqrt{\alpha} \sqrt{\pi}\right]}{\sqrt{\alpha}}.
\label{eq:f function approx}
\end{equation}
The XL-array consists of a large number of antenna elements, making \( N \) significantly large. Using the Taylor series expansion, when \(\alpha = \frac{N^2 d (1 - \theta^2)}{8r} \to \infty\), the function can be approximated as:
\begin{equation}
   2 \tilde{f}(\theta, r, \theta) = (-1)^{\frac{1}{4}} \sqrt{\frac{1}{\alpha}} + O\left(\frac{1}{\alpha}\right)^{\frac{3}{2}} .
\label{eq:f_theta_r_phi}
\end{equation}
Here, \((-1)^{\frac{1}{4}} \sqrt{\frac{1}{\alpha}}\), represents the leading contribution to the amplitude of \(f(\theta, r, \varphi)\), and thus the central beam gain can be expressed as
\begin{equation}
G\left(\mathbf{b}^H(\theta, r), \mathbf{a}(\theta)\right) \approx \frac{1}{2\sqrt{\alpha}},
\label{eq:G_theta_r}
\end{equation}

\subsection{Derivation of Beam Pattern Width}\label{subsec:Analysis of Beam Pattern Width}

 We define the normalized beam pattern gain as follows by normalizing the beam pattern using the central beam pattern gain defined in \Cref{subsec:Central Beam Pattern Gain}:
\begin{equation}
G_{\text{norm}}\left(\mathbf{h}^H(\theta, r), \mathbf{a}(\varphi)\right)=\frac{\lvert \mathbf{h}^H(\theta,r)\mathbf{a}(\varphi)\rvert}{\lvert \mathbf{h}^H(\theta,r)\mathbf{a}(\theta)\rvert}
    \label{eq:normalized G_2}
\end{equation}
According to \Cref{eq:channel model,eq:G_theta_r,eq:origin f,eq:G and f}, we rewrite \Cref{eq:normalized G_2} as
\begin{equation}
\begin{split}
&G_{\text{norm}}\left(\mathbf{h}^H(\theta, r), \mathbf{a}(\varphi)\right)\\
&=\frac{ G\left(\mathbf{b}^H(\theta,r),\mathbf{a}(\varphi)\right)}{G\left(\mathbf{b}^H(\theta,r),\mathbf{a}(\theta)\right)}\\
&=\frac{1}{2} \left|\operatorname{erf}\left(\frac{e^{\jmath\frac{3}{4} \pi}\sqrt{\pi}(\beta-2 \alpha) }{2 \sqrt{\alpha}}\right)-\operatorname{erf}\left(\frac{e^{\jmath \frac{3}{4} \pi}\sqrt{\pi}(\beta+2 \alpha) }{2 \sqrt{\alpha}}\right)\right|.
\end{split}
\label{eq:normalized G_final}
\end{equation}


We now analyze the properties of \Cref{eq:normalized G_final} with respect to $\beta$. Without loss of generality, we assume $\beta > 0$. First, when $\beta$ is small (i.e., $2\alpha-\beta\gg \sqrt{\alpha}$ and $2\alpha+\beta\gg \sqrt{\alpha}$), the following approximations hold: $\operatorname{erf}\left(\frac{e^{\jmath\frac{3}{4} \pi}\sqrt{\pi}(\beta-2 \alpha) }{2 \sqrt{\alpha}}\right)\to 1$ and $\operatorname{erf}\left(\frac{e^{\jmath\frac{3}{4} \pi}\sqrt{\pi}(\beta+2 \alpha) }{2 \sqrt{\alpha}}\right)\to -1$. In this case, the value of $G_{\text{norm}}\left(\mathbf{h}^{H}(\theta, r),\mathbf{a}(\varphi)\right)$ is comparable to $1$ and is approaching the central of the beam. We plot the normalized beam gain at $\alpha=6,\theta=0$ as an example, illustrated in the red region of the \Cref{fig:beta region}.This figure maps $\beta$ into $\varphi$ with $\varphi=\theta-\frac{2\beta}{N}$. And the small $\beta$ region is shaded by color red.

On the other hand, when $\beta$ is large, (i.e., $\beta-2\alpha\gg \sqrt{\alpha}$ and $\beta+2\alpha\gg \sqrt{\alpha}$),  $\operatorname{erf}\left(\frac{e^{\jmath\frac{3}{4} \pi}\sqrt{\pi}(\beta-2 \alpha) }{2 \sqrt{\alpha}}\right)\to -1$ and $\operatorname{erf}\left(\frac{e^{\jmath\frac{3}{4} \pi}\sqrt{\pi}(\beta+2 \alpha) }{2 \sqrt{\alpha}}\right)\to -1$ hold. As a result, the value of $G_{\text{norm}}\left(\mathbf{h}^{H}(\theta, r),\mathbf{a}(\varphi)\right)$ will approach zero. This large $\beta$ region is shown in \Cref{fig:beta region} in blue.

When $\beta$ has a moderate value (i.e., $2\alpha-\beta$ is comparable to $\sqrt{\alpha}$ and $\beta+2\alpha\gg \sqrt{\alpha}$), $G_{\text{norm}}\left(\mathbf{h}^{H}(\theta, r),\mathbf{a}(\varphi)\right)$ is approximated as
\begin{equation}
\tilde{G}_{\text{norm}}\left(\mathbf{h}^{H}(\theta, r),\mathbf{a}(\varphi)\right)=\frac{1}{2} \left|\operatorname{erf}\left(\frac{e^{\jmath\frac{3}{4} \pi}\sqrt{\pi}(\beta-2 \alpha) }{2 \sqrt{\alpha}}\right)+1\right|.
\label{eq:semi_approx of G norm}
\end{equation}
This moderate $\beta$ region is illustrated in green in \Cref{fig:beta region}. The relative error of this approximation is less than $4.8\%$ in the region $|\varphi|>1.95\times10^{-2}$, which means that the error quickly diminishes as the UE direction deviates from the array normal vector.
\begin{figure}[!t]
\captionsetup{justification=raggedright, singlelinecheck=false} 
\centering
\includegraphics[width=2.5in]{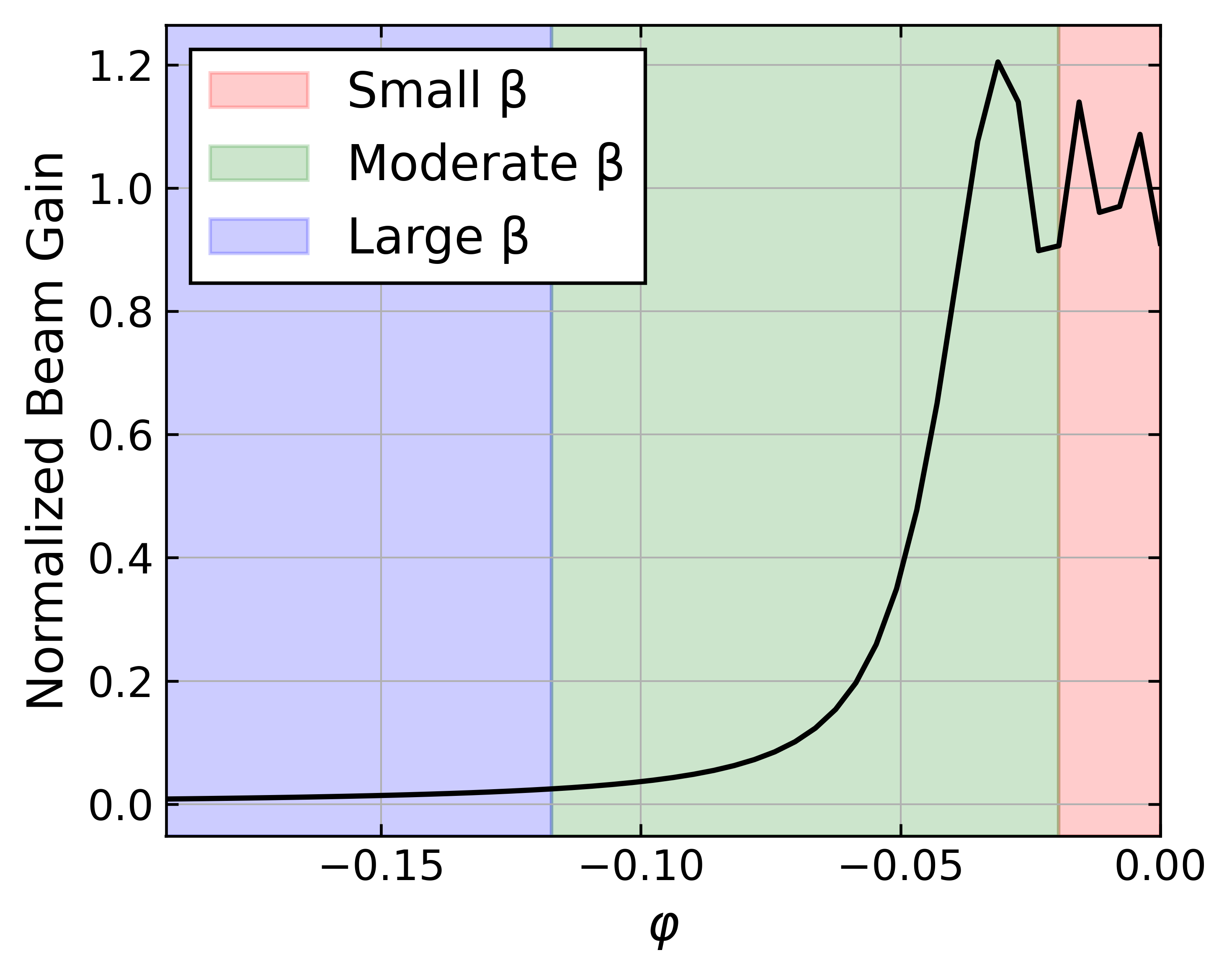}
\caption{Different Regions of $\beta$.}
\label{fig:beta region}
\end{figure}

We define the beam pattern width as a range of $\varphi$ similar to \cite{Interference_Perspective_for_Mixed_Near-_and_Far-Field}, within which the value of $G_{\text{norm}}\left(\mathbf{b}^{H}(\theta, r),  \mathbf{a}(\varphi) \right) $ is relatively big with certain user at a position of $(\theta, r)$. To be specific, we define following beam pattern width
\begin{equation}
    \Phi^\rho = \{\varphi \mid G_{\text{norm}}\left(\mathbf{h}^{H}(\theta, r),\mathbf{a}(\varphi)\right) > \rho\},
    \label{Main angle set for G}
\end{equation}
\begin{equation}
    B(\theta,r) = \operatorname{Range}(\Phi^\rho) ,
    \label{eq:simulated width}
\end{equation}
where $\Phi^\rho$ is the set of corresponding $\varphi$ at which $G_{\text{norm}}\left(\mathbf{h}^{H}(\theta, r),\mathbf{a}(\varphi)\right)$ has larger value than threshold $\rho$ and $\rho$ is a specific value of $G_{\text{norm}}\left(\mathbf{h}^{H}(\theta, r),\mathbf{a}(\varphi)\right)$ within moderate $\beta$ region. We call the angle set like this \textbf{Main Angle Set} (MAS).

Let $s=\frac{\sqrt{\pi}(\beta-2 \alpha)}{2 \sqrt{\alpha}}$, the plot of $\tilde{G}_{\text{norm}}\left(s\right)$ in \Cref{eq:semi_approx of G norm} is shown in \Cref{fig:f_3}. The green region is valid for approximation of green region in \Cref{fig:beta region}.
\begin{figure}[!t]
\captionsetup{justification=raggedright, singlelinecheck=false} 
\centering
\includegraphics[width=2.5in]{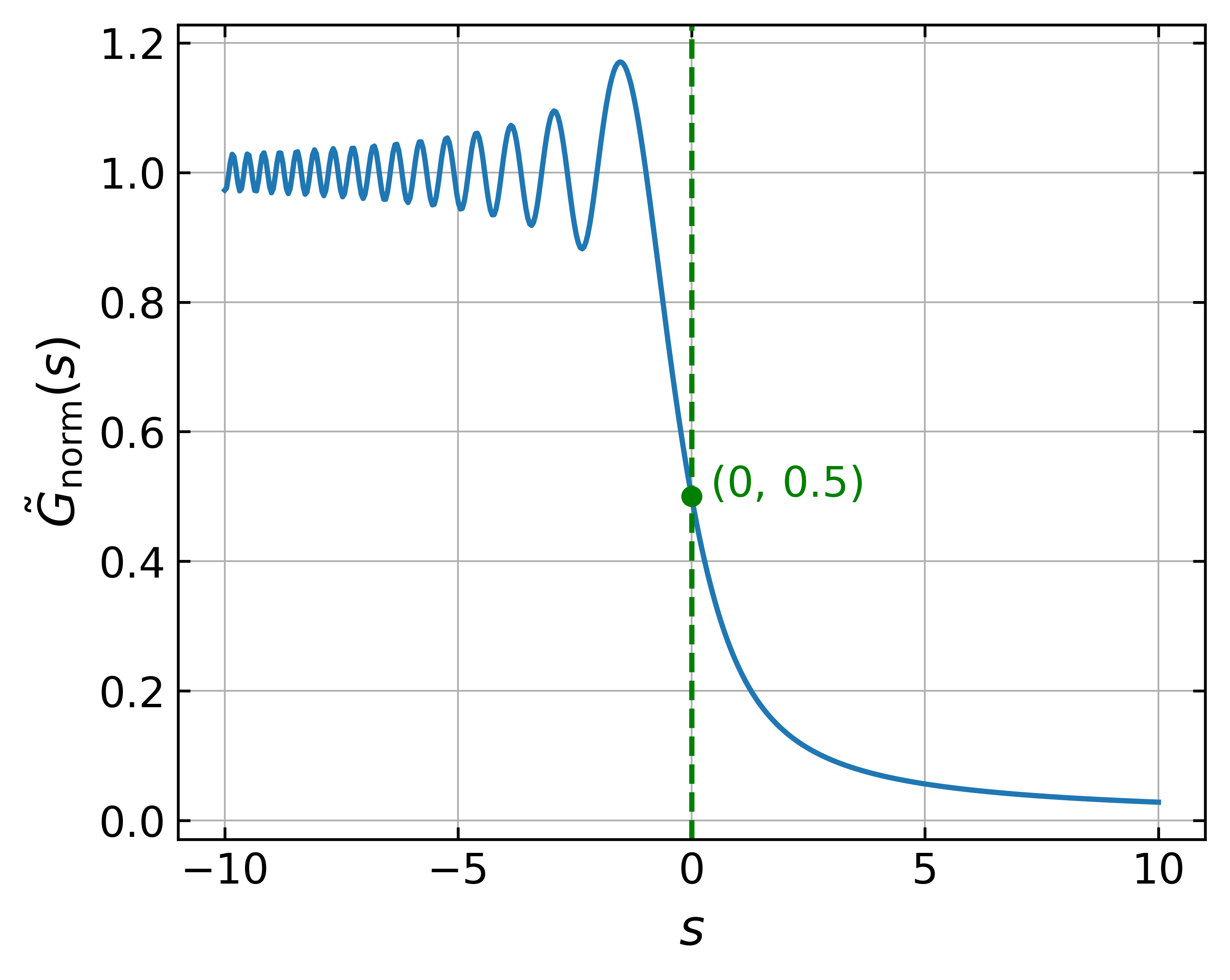}
\caption{Approximation of beam gain.}
\label{fig:f_3}
\end{figure}
From this plot we can get the injective mapping of $s$ and $\rho$. Given any $\rho$, the corresponding $s_\rho$ is given by
\begin{equation}
    \rho=\frac{1}{2}\lvert \operatorname{erf}(e^{\jmath \frac{3}{4} \pi}s_\rho)+1\rvert
    \label{eq:threshold rho wrt s rho}
\end{equation}
Namely, when $\beta>0$, $G_{\text{norm}}\left(\mathbf{h}^{H}(\theta, r),\mathbf{a}(\varphi)\right)>\rho$ is satisfied when 
\begin{equation}
  \beta<2\alpha+\frac{2s_\rho}{\sqrt{\pi}}\sqrt{\alpha}  
\end{equation}
From $\alpha=\frac{N^{2} d\left(1-\theta^{2}\right)}{8 r}$, $\beta=\frac{N(\theta-\varphi)}{2}$ and the symmetric of $G_{\text{norm}}\left(\mathbf{h}^{H}(\theta, r),\mathbf{a}(\varphi)\right)$ w.r.t. $\theta-\varphi$, one can obtain the beam pattern width as a function of $\theta$ and $r$:
\begin{equation}
    B^\rho(\theta,r)=\frac{Nd(1-\theta^2)}{r}+\frac{4s_\rho}{\sqrt{\pi}}\sqrt{\frac{d(1-\theta^2)}{2r}}.
    \label{eq:close_form of B}
\end{equation}
An example is shown in \Cref{fig:f_3} in green point. When $\rho=0.5$, the corresponding $s_\rho$ is $0$, and beam pattern width is
\begin{equation}
    B^{(0.5)}(\theta,r)=\frac{Nd(1-\theta^2)}{r}.
    \label{B at 0.5}
\end{equation}
We will use \Cref{B at 0.5} to guide our following beam training design. 
\subsection{Derivation of Modified Rayleigh distance}
The conventional Rayleigh distance is based on the criterion that the maximum phase error for each antenna element does not exceed $\frac{\pi}{8}$. However, these boundaries do not account for communication performance or beam management losses. Therefore, we define a modified Rayleigh distance for beam training based on the normalized beam pattern $G_{\text{norm}}\left(\mathbf{h}^{H}(\theta, r),\mathbf{a}(\varphi)\right)$ to better distinguish NUs and FUs. 

For ULA of $N$ antenna elements, the resolution of far-field beam pattern \cite{Tutorial_with_beam_pattern_and_resolution} will be $\frac{2}{N}$. We define the modified Rayleigh distance based on the fact that, using DFT codebook, the beam width $B^{\rho}(\theta,r)$ should be larger than a threshold  $\Delta_p=p\cdot\frac{2}{N}$, where $p$ is a small constant denoted as beam width scaling factor. A larger $p$ value permits a broader beam width, resulting in a smaller modified Rayleigh distance, and vice versa. Thus, at certain $\theta$, modified Rayleigh distance $R^{\rho,p}_{mRay}$ will satisfy following equation
\begin{equation}
     B^{\rho}(\theta,R^{\rho,p}_{mRay})=\Delta_p.
     \label{B_rayleigh_p}
\end{equation} 
Combing \Cref{eq:close_form of B} and \Cref{B_rayleigh_p}, we obtain modified Rayleigh distance for any $\theta$ as
\begin{equation}
\begin{split}
R^{\rho,p}_{mRay}(\theta) = \Biggl(\frac{N}{4p}\Biggl[ & \frac{4s_\rho}{\sqrt{\pi}}\sqrt{\frac{d(1-\theta^2)}{2}} \\[1mm]
 & + \sqrt{8d(1-\theta^2)\Bigl(\frac{(s_\rho)^2}{\pi}+p\Bigr)} \Biggr] \Biggr)^2.
\end{split}
\label{Rmray_close}
\end{equation}
For example, when $\rho=0.5,p=3$, the closed-form of modified Rayleigh distance is
\begin{equation}
    R^{(0.5,3)}_{mRay}(\theta)=\frac{N^2d(1-\theta^2)}{6}.
    \label{Rayleigh_theoretical}
\end{equation}
This indicates that the modified Rayleigh distance is angle-dependent and decreases with smaller $\theta$.  Moreover, it is significantly smaller than the conventional Rayleigh distance.  For example, the modified Rayleigh distance is only $\frac{1}{6}$ of conventional Rayleigh distance at $\theta=0$.
\subsection{The Verification of Beam Pattern Width}

In this section, we verify the derived beam pattern width in \Cref{B at 0.5} and modified Rayleigh distance in \Cref{Rayleigh_theoretical}. In this paper, we set $N=512$ , carrier frequency $f_c=100\text{GHz}$ and $d=\frac{\lambda}{2}=1.5$mm. 

\subsubsection{Beam Width}We set $\theta=0, r\in[5\text{m},70\text{m}]$, which is within the modified Rayleigh distance and generate normalized beam gain within this region. For each gain we calculate the simulated beam width as in \Cref{eq:simulated width} and theoretical beam gain as \Cref{eq:close_form of B}. \Cref{Beam width at fixed angle} shows both simulated and theoretical beam width as a function of distance when \(\theta\) is fixed at zero. The dashed line in the figure is the linear fit of the simulated beam width and the $1/r$. Following the theoretical analysis, the  beam width \(B\) is expected to be linearly depend on  \(1/r\) with a slope of \(Nd = 0.768\). The simulation results confirm this linear pattern, yielding a slope of \(0.771\). Then, the distance is fixed at \(r=5\text{m}\) while \(\theta\) varies in the range \([-0.97,0.97]\) to verify the relationship between \(B\) and \(1-\theta^2\). According to the theoretical analysis, \(B\) should be linearly dependent on \(1-\theta^2\) with a slope of \(\frac{Nd}{r}\approx 0.1536\). The simulated results align with this prediction, showing a slope of \(0.1542\), as illustrated in  \Cref{Beam width at fixed distance}. The beam width exhibits a step-like increase due to finite antenna resolution, which forces the measured beam width to be an integer multiple of the sampling interval \(\tfrac{2}{N}\).
\begin{figure}[!t]
\captionsetup{justification=centering, singlelinecheck=true} 
\centering
\subfloat[\tiny(a)][\textrm{\small Beam width versus distance at angle $\theta=0$}]{%
    \includegraphics[width=2.5in]{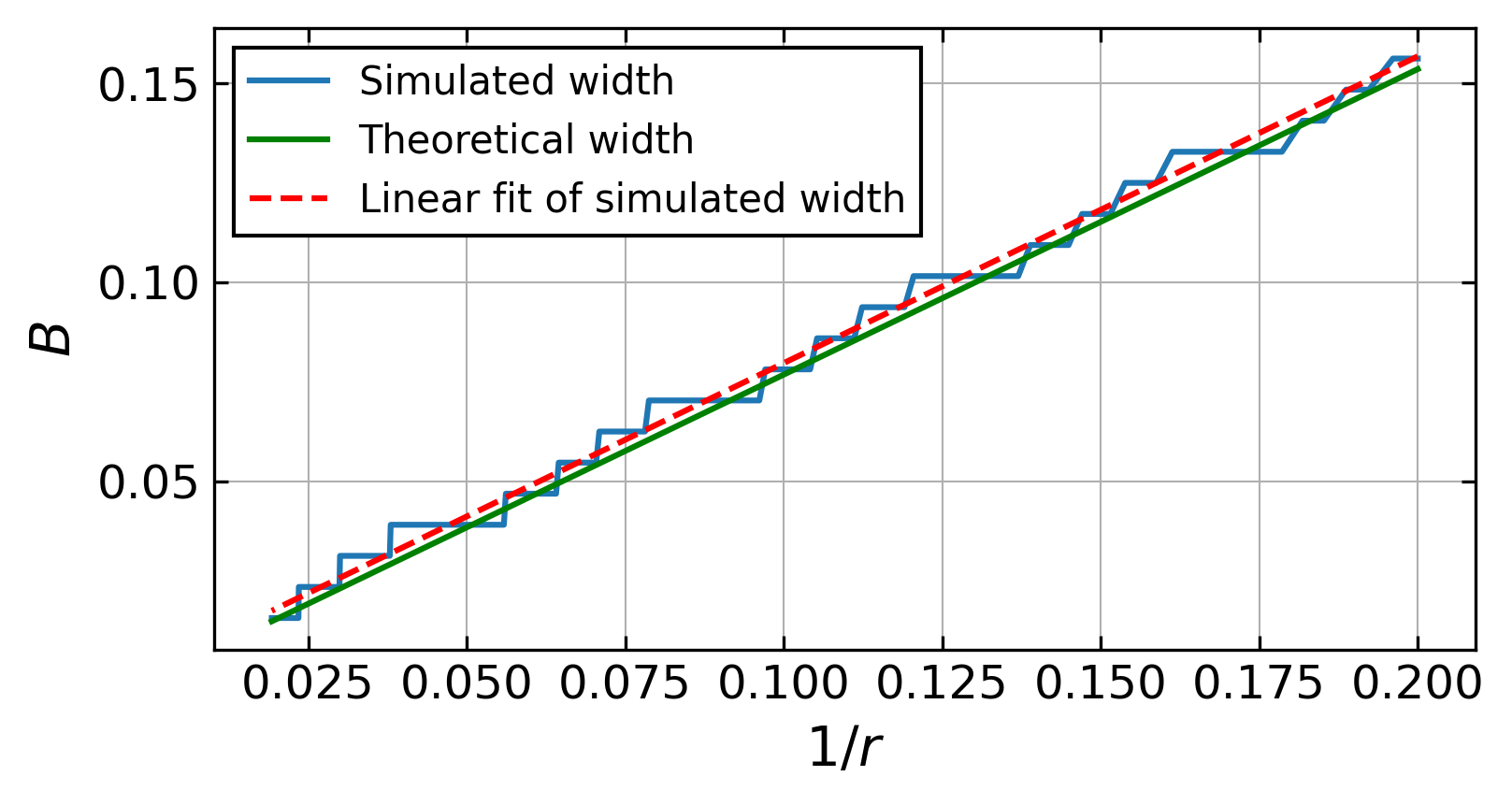}%
    \label{Beam width at fixed angle}
}\\
\subfloat[\tiny(b)][\textrm{\small Beam width versus angle at distance $r=5$m}]{%
    \includegraphics[width=2.5in]{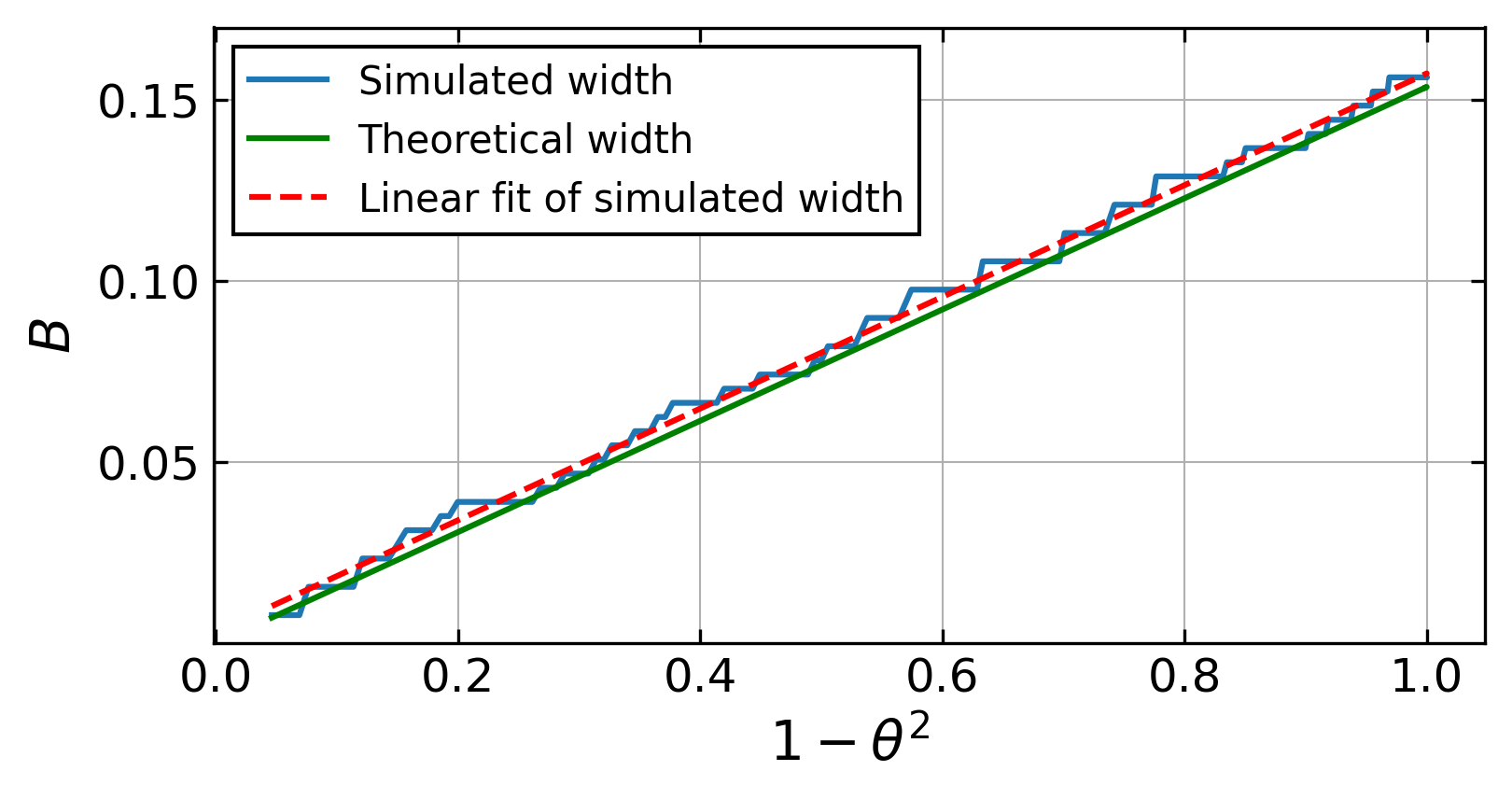}%
    \label{Beam width at fixed distance}
}
\captionsetup{justification=raggedright, singlelinecheck=false} 
\caption{Validation of beam width within Rayleigh distance. }
\label{Validation of beam width in theta and r.}
\end{figure}

\subsubsection{Modified Rayleigh distance}We generate the normalized beam pattern of $G_{\text{norm}}\left(\mathbf{h}^{H}(\theta, r),\mathbf{a}(\varphi)\right)$ at different angles $\theta_n$, and the smallest distance $r$ that satisfies \Cref{B_rayleigh_p} at $\rho=0.5,p=3$ is the simulated modified Rayleigh distance $R^{0.5,3}_{mRay}(\theta_n)$ for certain $\theta_n$. 
\Cref{fig:Rayleigh distance at different angle} compares the simulated modified Rayleigh distance and the theoretical modified Rayleigh distance in \Cref{Rayleigh_theoretical}. 
It can be seen that $R_{mRay}$ and $1-\theta^2$ demonstrate a linear correlation. This is consistent with theoretical derivation in \Cref{Rayleigh_theoretical}. The slope is $65.54$, which is close to $\frac{N^2d}{6}=66.73$.
\begin{figure}[!t]
\centering
\includegraphics[width=2.5in]{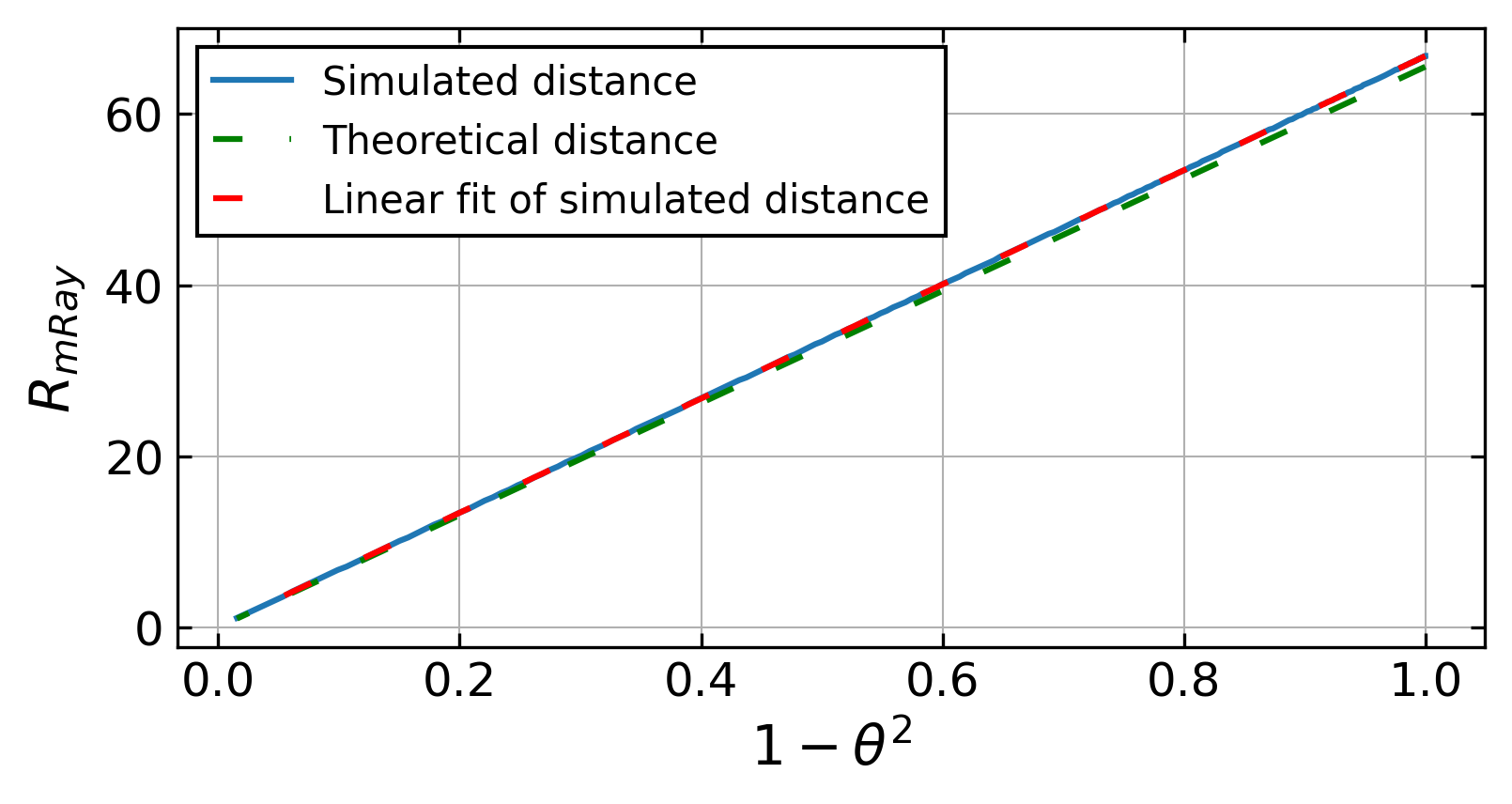}
\captionsetup{justification=raggedright, singlelinecheck=false}
\caption{Rayleigh distance versus angle.}
\label{fig:Rayleigh distance at different angle}
\end{figure}

\section{Low-complexity Scheme}\label{sec:coarse and refine}
Based on the close-form results, we propose a low-complexity near-field beam training scheme by jointly estimating user angle and distance.
\subsection{Estimate Angle by Middle Theta}\label{subsec:Estimate Angle by Middle Theta}
For a user located at  $(\theta,r)$, BS first performs beam sweeping using the DFT codebook to obtain the receive signal amplitude
\begin{equation}
    \{|y(\mathbf{v}_n)|\mid y(\mathbf{v}_n)=\mathbf{h}^{H}\left(\theta, r\right)\mathbf{v}_n x+w,\mathbf{v}_n\in\mathbf{V_{DFT}}\}.
\end{equation}

We adopt and improve the method in \cite{Fast_Near-Field_Beam_Training}, which estimates the user angle $\theta$ by the MAS of $\{|y(\mathbf{v_n})|\}$. We improve the median theta method, especially in low SNR conditions by introducing a clustering step that groups together indexes of strong beam pattern gains. 

Define the set of candidate indices as
\begin{equation}
\mathcal{I} = \{ n \mid |y(\mathbf{v}_n)| > \rho_2 \}.
\end{equation}
Next, we partition \(\mathcal{I}\) into clusters based on proximity. Specifically, two consecutive indices \(n\) and \(n'\) belong to the same cluster if \(n' - n \le L\),where \(L\) is a fixed small number. 
If \(L\) is too large, peaks due to noise variations may be grouped with the main-beam peaks, contaminating the cluster and degrading the angle estimation. Conversely, if \(L\) is too small, adjacent peaks  may be split into separate clusters. In our algorithm, we choose \(L=8\). The detailed algorithm is in \Cref{alg:clustering}.

After grouping the indices into clusters, we examine each cluster by finding the strongest received signal within it. The cluster whose strongest signal is higher than those in all other clusters is selected as the optimal cluster $\mathcal{C}^*$. 
The MAS is 
\begin{equation}
\Phi^{\rho_2} = \{\varphi_n \mid n \in \mathcal{C}^*\}.
\end{equation}
Here we choose $\rho_2=0.65\operatorname{max}(\{y(\mathbf{v}_n)\})$. This is a different value from \cite{Fast_Near-Field_Beam_Training}, however, this threshold will bring increase in performance at low SNR scenario, and thus we use $\rho_2$ for fair comparison.Finally, we estimate the user angle as the midpoint of \(\Phi^{\rho_2}\):
\begin{equation}
\hat{\theta}_0 = \frac{\max(\Phi^{\rho_2})+\min(\Phi^{\rho_2})}{2}.
\end{equation}
A median-k-selection scheme is also proposed in \cite{Fast_Near-Field_Beam_Training}, which choose k angles $\hat{\theta}_1,\hat{\theta}_2,\cdots, \hat{\theta}_k$ whose index are in the vicinity of the median angle $\hat{\theta}_0$ in $\Phi^{\rho_2}$ for further estimation. 
\begin{algorithm}[!h]
\caption{Beam Sweeping Results Clustering}
\label{alg:clustering}
\begin{algorithmic}[1]
\REQUIRE Beam gains $\{|y(\mathbf{v}_n)|\}_{n=1}^{N}$, threshold $\rho_2$, gap $L$.
\ENSURE Clusters $\mathcal{S}=\{\mathcal{C}_1,\dots,\mathcal{C}_M\}$.
\STATE Initialize $\mathcal{C} \gets \varnothing$, $\mathcal{S} \gets \varnothing$, $n_{\text{last}} \gets -1$.
\FOR{$n=1$ \TO $N$}
    \IF{$|y(\mathbf{v}_n)| > \rho_2$}
        \IF{$\mathcal{C}=\varnothing$ \OR $(n - n_{\text{last}} \le L)$}
            \STATE $\mathcal{C} \gets \mathcal{C}\cup\{n\}$; $n_{\text{last}} \gets n$.
        \ELSE
            \STATE Append $\mathcal{C}$ to $\mathcal{S}$; set $\mathcal{C} \gets \{n\}$; $n_{\text{last}} \gets n$.
        \ENDIF
    \ENDIF
\ENDFOR
\IF{$\mathcal{C} \neq \varnothing$} \STATE Append $\mathcal{C}$ to $\mathcal{S}$.
\ENDIF
\RETURN $\mathcal{S}$.
\end{algorithmic}
\end{algorithm}
\subsection{Estimate Distance by Beam Width}\label{subsec:estimate_distance}
With the estimated angle $\hat{\theta}$ in Section IV.B, the beam width of the received signal is defined as follows
\begin{equation}
\begin{split}
   & \hat{\Phi}^\rho = \left\{\varphi_n \mid \left|\frac{y(\mathbf{v}_n)}{y(\mathbf{a}(\hat{\theta}))}\right| > \rho\right\},\\
    &\hat{B}^\rho = \operatorname{Range}(\hat{\Phi}^\rho) .
\end{split}
\label{eq:beam width}
\end{equation}
where ${y(\mathbf{a}(\hat{\theta}))}$ is the received signal when BS is using codeword $\mathbf{a}(\hat{\theta})$.  

We consider $\left|\frac{y(\mathbf{v}_n)}{y(\mathbf{a}(\hat{\theta}))}\right|=\hat{G}_{\text{norm}}\left(\mathbf{h}^H(\theta, r), \mathbf{a}(\varphi_n)\right)$ is the sampled estimation of $G_{\text{norm}}\left(\mathbf{h}^H(\theta, r), \mathbf{a}(\varphi)\right)$. $\hat{B}^\rho$ is the estimated beam width for user $(\theta,r)$. If $\hat{B}^\rho\le\Delta_\theta$, the user is considered as FU, and DFT codeword $\mathbf{a}(\hat{\theta})$ will be used for beamforming. Otherwise, we classify the user as NU.  By solving \Cref{eq:close_form of B} the distance of this user is estimated as
\begin{equation}
\hat{r}^\rho
= \frac{d\bigl(1-\hat{\theta}^2\bigr)}{\bigl(\hat{B}^\rho\bigr)^2}
\left(
\sqrt{\frac{2}{\pi}}\,s_\rho
\;+\;
\sqrt{
  \frac{2\bigl(s_\rho\bigr)^2}{\pi}
  \;+\;
  \hat{B}^\rho\,N
}
\right)^{2}.
\label{eq:estimated_r_rho}
\end{equation}
Then near-field codeword $\mathbf{b}(\hat{\theta},\hat{r}^\rho)$ is adopted. If a median-k-selection is used for estimation of $\theta$, we will find the estimated $\hat{r}_i^\rho$ for all k $\hat{\theta}_i$ and perform additional beam refinement using k $\mathbf{b}(\hat{\theta}_i,\hat{r}_i^\rho)$. The codeword with the highest received power will finally be adopted. The detailed algorithm is in \Cref{alg:proposed}.
\begin{algorithm}[!h]
\caption{Coarse Near-Field Beam Training Scheme}
\label{alg:proposed}
\begin{algorithmic}[1]
\REQUIRE DFT codebook \(\mathbf{V}_{\mathrm{DFT}}\); parameters \((N,d,\Delta_\theta,\rho,L,k)\).
\ENSURE Estimated angle \(\hat{\theta}\), distance \(\hat{r}^\rho\), and beamforming codeword.
\STATE \textbf{Angle Estimation:}
\begin{enumerate}
    \item Perform beam sweeping to obtain \(\{|y(\mathbf{v}_n)|\}\) and threshold \(\rho_2\).
    \item Cluster the indices using Algorithm~\ref{alg:clustering} with $L$ and select the cluster \(\mathcal{C}^*\) with the highest peak amplitude.
    \item Form the MAS \(\Phi^{\rho_2}=\{\varphi_n \mid n\in \mathcal{C}^*\}\).
    \item Estimate \(\hat{\theta}_0\) as the median of \(\Phi^{\rho_2}\) and choose \(k\) candidate angles \(\hat{\theta}_1,\hat{\theta}_2,\dots,\hat{\theta}_k\) near \(\hat{\theta}_0\).
\end{enumerate}
\STATE \textbf{Distance Estimation:}
\begin{enumerate}
    \item \textbf{For} \(i=1\) \textbf{to} \(k\):
    \begin{enumerate}
         \item normalize the beam gain for candidate $\hat{\theta}_i$: compute $\dfrac{y(\mathbf{v}_n)}{y(\mathbf{a}(\hat{\theta}_i))}$.
         \item Find MAS \(\hat{\Phi}_i^\rho=\Bigl\{\varphi_n : \Bigl|\dfrac{y(\mathbf{v}_n)}{y(\mathbf{a}(\hat{\theta}_i))}\Bigr|>\rho\Bigr\}\) and let \(\hat{B}_i^\rho=\operatorname{Range}(\hat{\Phi}_i^\rho)\).
        \item \textbf{If} \(\hat{B}_i^\rho \le \Delta_\theta\) \textbf{then} set \(\hat{r}^\rho_i \gets \infty\) and \(\mathbf{v}_i \gets \mathbf{a}(\hat{\theta}_i)\); \textbf{else} set \(\hat{r}^\rho_i\) via Eq.~(\ref{eq:estimated_r_rho}) and \(\mathbf{v}_i \gets \mathbf{b}(\hat{\theta}_i,\hat{r}^\rho_i)\).

    \end{enumerate}
    \item Select the candidate \((\hat{\theta},\hat{r}^\rho)\) with the highest beamforming response.
\end{enumerate}
\RETURN \((\hat{\theta},\hat{r}^\rho)\) and \(\mathbf{v}^*\).
\end{algorithmic}
\end{algorithm}

\subsection{Refine Estimation of Distance by Iteration}\label{sec:refine of beam width}

The beam pattern width defined above is based on the assumption that  $\alpha\to\infty$. We can further refine the results of the coarse scheme and increase the estimation accuracy using the exact expression iteratively.

When $\beta=2\alpha+\frac{2s_\rho}{\sqrt{\pi}}\sqrt{\alpha}$, we denote $\varphi$ at this time as $\varphi_\rho$. And beam gain at $\varphi_\rho$ is $G\Bigl(\mathbf{b}^H(\theta, r),\mathbf{a}(\varphi_\rho)\Bigr)$.
From \Cref{eq:f function approx,eq:G and f}, the central beam height without Taylor approximation is
\begin{equation}
        G\Bigl(\mathbf{b}^H(\theta, r),\mathbf{a}(\theta)\Bigr)=  \frac{1}{2}\left|\frac{ \, \mathrm{erf}\left[e^{\jmath\frac{3}{4}\pi} \sqrt{\alpha} \sqrt{\pi}\right]}{\sqrt{\alpha}}\right|.
\end{equation}
Thus from \Cref{eq:normalized G_final}, the exact threshold when using $\rho$ and corresponding $s_\rho$ is
\begin{equation}
\begin{split}
    &\rho_{\text{exact}} \\
    & =\frac{G\Bigl(\mathbf{b}^H(\theta, r),\mathbf{a}(\varphi_\rho)\Bigr)}{ G\Bigl(\mathbf{b}^H(\theta, r),\mathbf{a}(\theta)\Bigr)}\\
   &= \frac{1}{2} \cdot \Biggl| \frac{\operatorname{erf}\Bigl(e^{\jmath\frac{3}{4}\pi}\, s_\rho\Bigr) - \operatorname{erf}\Bigl(e^{\jmath\frac{3}{4}\pi}\Bigl(2\sqrt{\alpha\pi}+ s_\rho\Bigr)\Bigr)}{\operatorname{erf}\Bigl(e^{\jmath\frac{3}{4}\pi} \sqrt{\pi\,\alpha}\Bigr)} \Biggr|. 
\end{split}
\label{exact_threshold}
\end{equation}

Based on this, we propose an iterative-method-based refined scheme to improve the accuracy of beam training. In each iteration, using estimated angle $\hat{\theta}$ and distance $\hat{r}^\rho$ in previous iteration, we calculate the estimated $\hat{\alpha}$ as
\begin{equation}
    \hat{\alpha}=\frac{N^2d(1-\hat{\theta}^2)}{8\hat{r}^\rho}.
    \label{eq:est_alpha}
\end{equation}
We then get estimated threshold $\hat{\rho}_{\text{exact}}$ and continue the distance estimation step in \Cref{subsec:estimate_distance} to get new $\hat{\alpha}$. The iteration is performed until the same beam pattern width as the last iteration is obtained. The detailed algorithm is as in \Cref{alg:enhanced}.
\begin{algorithm}[!h]
\caption{Refined Beam Training Scheme}
\label{alg:enhanced}
\begin{algorithmic}[1]
\REQUIRE Initial estimates \((\hat{\theta},\hat{r}_0)\), initial beam width \(B_0^\rho\), parameters \((N,d,\rho)\), tolerance \(\varepsilon\), and maximum iterations \(I_{\text{max}}\).
\ENSURE Refined distance \(\hat{r}^{\rho}\) and codeword \(\mathbf{b}(\hat{\theta},\hat{r}^{\rho})\).
\STATE \textbf{Initialize:} 
\[
\hat{\alpha}_0\gets\frac{N^2d(1-\hat{\theta}^2)}{8\hat{r}_0},\hat{\rho}_{\mathrm{exact,0}} \text{ (using $\hat{\alpha}_0$ in Eq.~\ref{exact_threshold})}, t\gets 1.
\]
\REPEAT
    \STATE Measure beam width \(B_t^{\rho_{\text{exact}}}\)
    using new MAS with \(\hat{\rho}_{\mathrm{exact},{t-1}}\).
    \STATE Update distance: \(\hat{r}_t^{\rho_{\text{exact}}}\) is obtained via Eq.~(\ref{eq:estimated_r_rho}) with \(B^\rho = B_t^{\rho_{\text{exact}}}\).
    \STATE Update \(\hat{\alpha}_t\gets\frac{N^2d(1-\hat{\theta}^2)}{8\hat{r}_t^{\rho_{\text{exact}}}}\) and obtain \(\hat{\rho}_{\mathrm{exact},t}\) via \Cref{exact_threshold}.
    \STATE Increment \(t\gets t+1\).
\UNTIL{\(\bigl|B_t^{\rho_{\text{exact}}} - B_{t-1}^{\rho_{\text{exact}}}\bigr|<\varepsilon\) or \(t\ge I_{\text{max}}\)}
\RETURN \(\hat{r}^\rho=\hat{r}_{t-1}^{\rho_{\text{exact}}}\) and \(\mathbf{v}^*=\mathbf{b}(\hat{\theta},\hat{r}^\rho)\).
\end{algorithmic}
\end{algorithm}
\subsection{Performance Evaluation}
In this section, we will present the results of simulation and validate the performance of the coarse and refined scheme. We set the reference SNR as the SNR of a user at $(0, 5\text{m})$ without beamforming. Transmitted power is fixed. We first test the influence of different k using the median-k-selection scheme on the proposed two approach. 
In a single-user scenario, the BS utilizes the beamforming vector $\mathbf{v}^*=\mathbf{b}(\hat{\theta},\hat{r}^\rho)$, which is determined using the estimated angle $\hat{\theta}$ and distance $\hat{r}^\rho$.  The achievable rate is defined as $R=\operatorname{log_2}(1+\frac{|\mathbf{h}^{H}\left(\theta, r\right)\mathbf{v}|^2}{\sigma^2})$. SNR is set as $6$ dB. And the result can be found in \Cref{achi_k}. It can be seen that $k=3$ has a much better achievable rate than $k=1$. When $k>3$, the improvement of achievable rate by increasing $k$ is minor. Thus for the sake of simplicity, we use $k=3$ for simulation in the rest of our analysis.
\begin{figure}[!t]
\captionsetup{justification=raggedright, singlelinecheck=false} 
\centering
\includegraphics[width=2.5in]{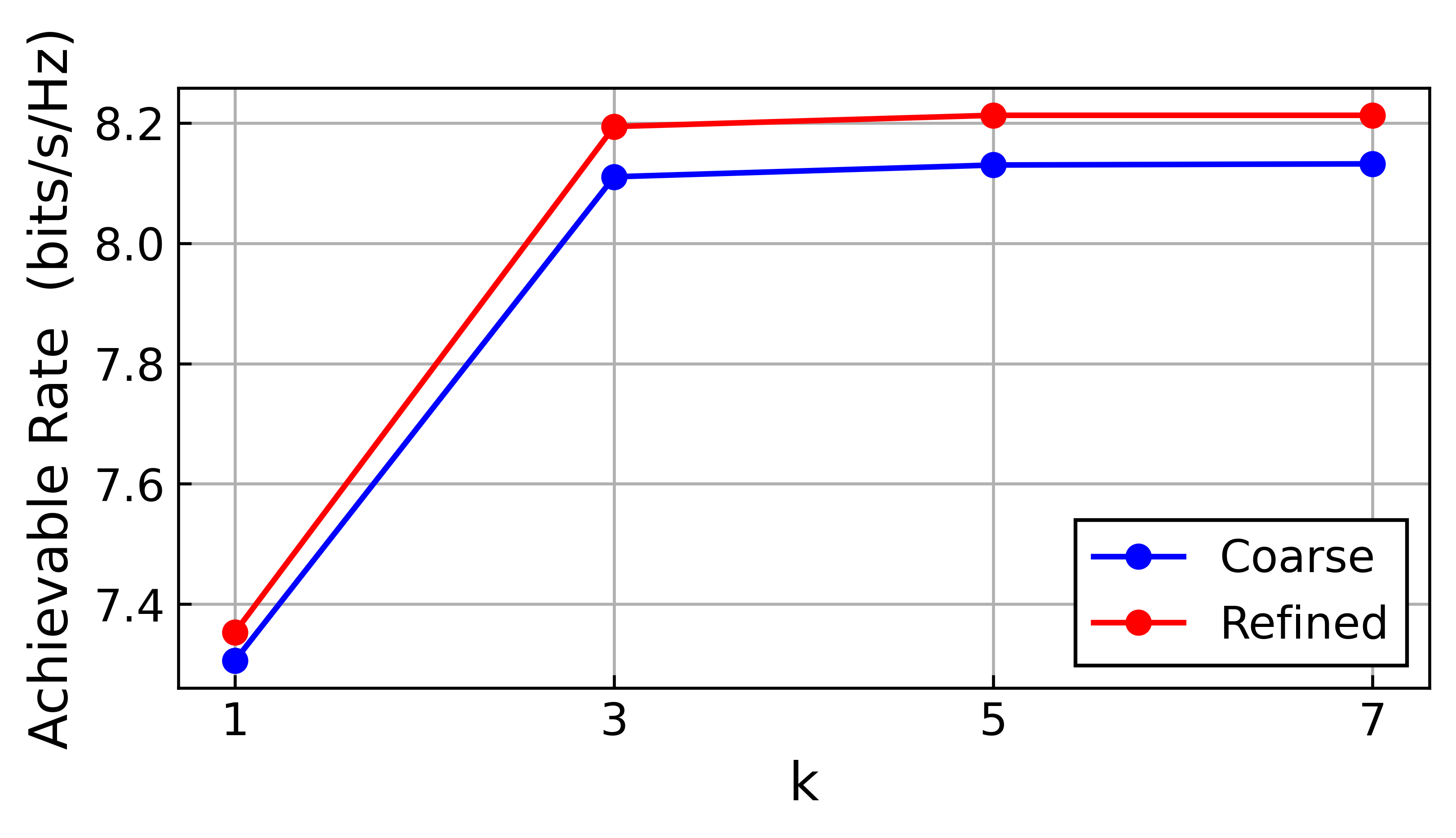}
\caption{Achievable rate versus reference SNR for different k.}
\label{achi_k}
\end{figure}
We sweep different noise power by setting reference SNR from 4 dB to 30 dB. We set the user locations within the modified Rayleigh distance and thus all users are NUs. 

\subsubsection{Accuracy}Firstly, we test the accuracy of coarse scheme and refined scheme. Mean Square Error (MSE) is used for metrics. In our simulation, 1000 NUs with random positions \((\theta, r)\) are generated. 
We compare our approach with the following schemes:
\begin{enumerate}
\item{Joint angle and distance estimation (Joint scheme).} This is proposed  in \cite{Joint_Angle_and_Range_Estimation}. We set  $k=3$ for a fair comparison.
\item{Fast estimation using DFT code (Fast scheme).} This is proposed in \cite{Fast_Near-Field_Beam_Training}. We also consider $k=3$.
\item{Exhausted beam training (Exhausted scheme).} This is to use the near-field codebook proposed in \cite{Cui_and_Dai_channel_model}. We set $\beta=1.6$, which is the same as in \cite{Cui_and_Dai_channel_model}.
\end{enumerate} 
The results can be found in \Cref{Angle MSE,Distance MSE}. Our proposed two schemes achieve the lowest MSE in both angle and distance estimation. For angle estimation, as seen in \Cref{Angle MSE}, the clustering method enables lower MSE than other schemes, in low SNR regions. For distance estimation as shown in \Cref{Distance MSE}, the proposed scheme consistently outperforms the others, especially at higher SNR levels, with MSE improvements of up to one order of magnitude. The refined scheme perform worse than coarse scheme in low SNR but better in high SNR because the iterative method rely heavily on the accuracy of beam pattern. The exhausted scheme performs worse than proposed two schemes because the resolution of it is limited. In low SNR, the beam pattern is severely influenced by noise, and thus the refined results from iteration is not precise. The inaccuracy in distance estimation also influences the accuracy in angle estimation in low SNR for refined scheme because for $k=3$, the angle and distance is jointly estimated by using beamforming as evaluation.

\begin{figure}[!t]
\captionsetup{justification=raggedright, singlelinecheck=false} 
\centering
\subfloat[\tiny(a)][\textrm{\small MSE of angle versus reference SNR.}]{%
    \includegraphics[width=0.5\linewidth]{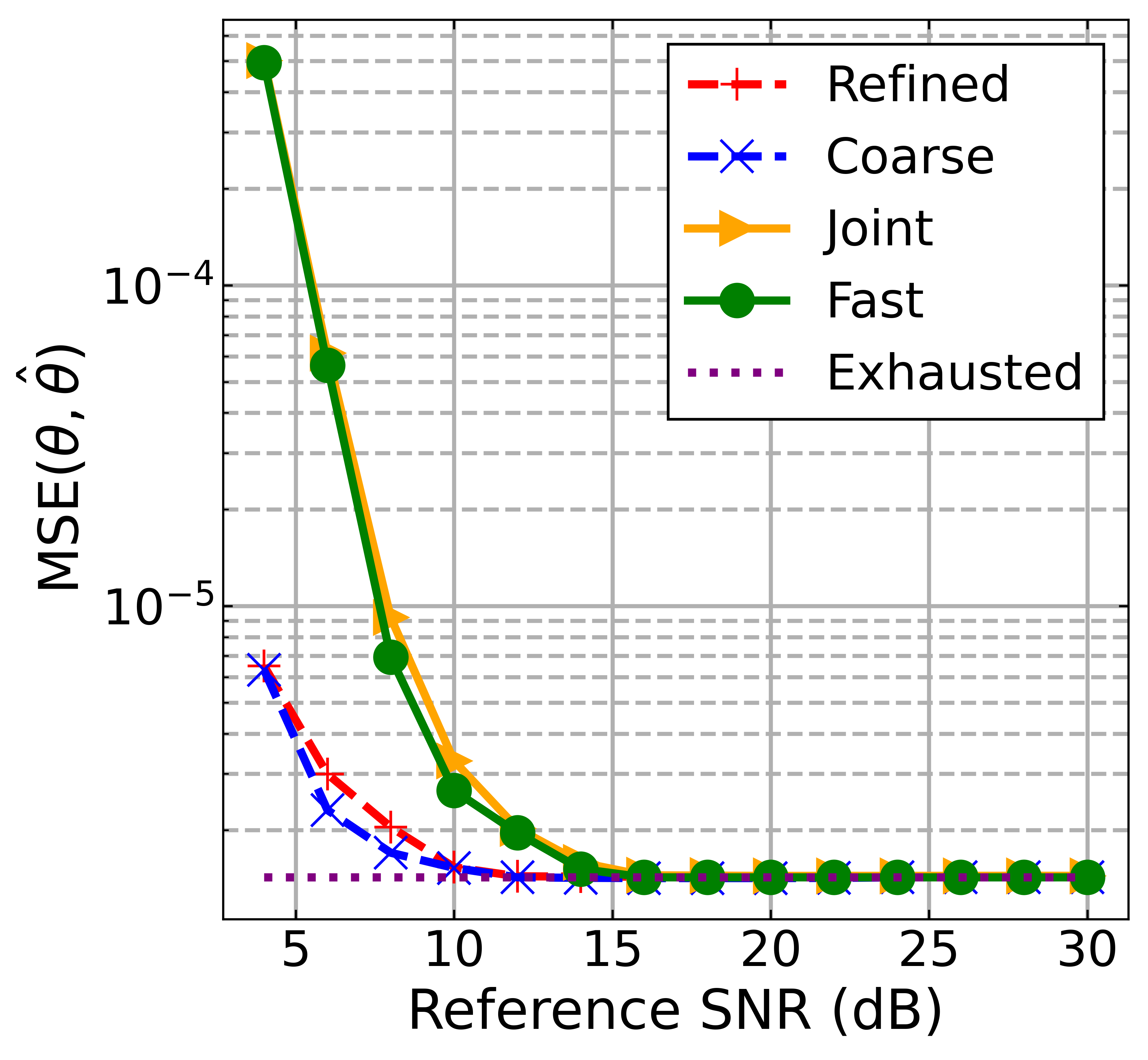}%
    \label{Angle MSE}%
}%
\subfloat[\tiny(a)][\textrm{\small MSE of distance versus reference SNR.}]{%
    \includegraphics[width=0.5\linewidth]{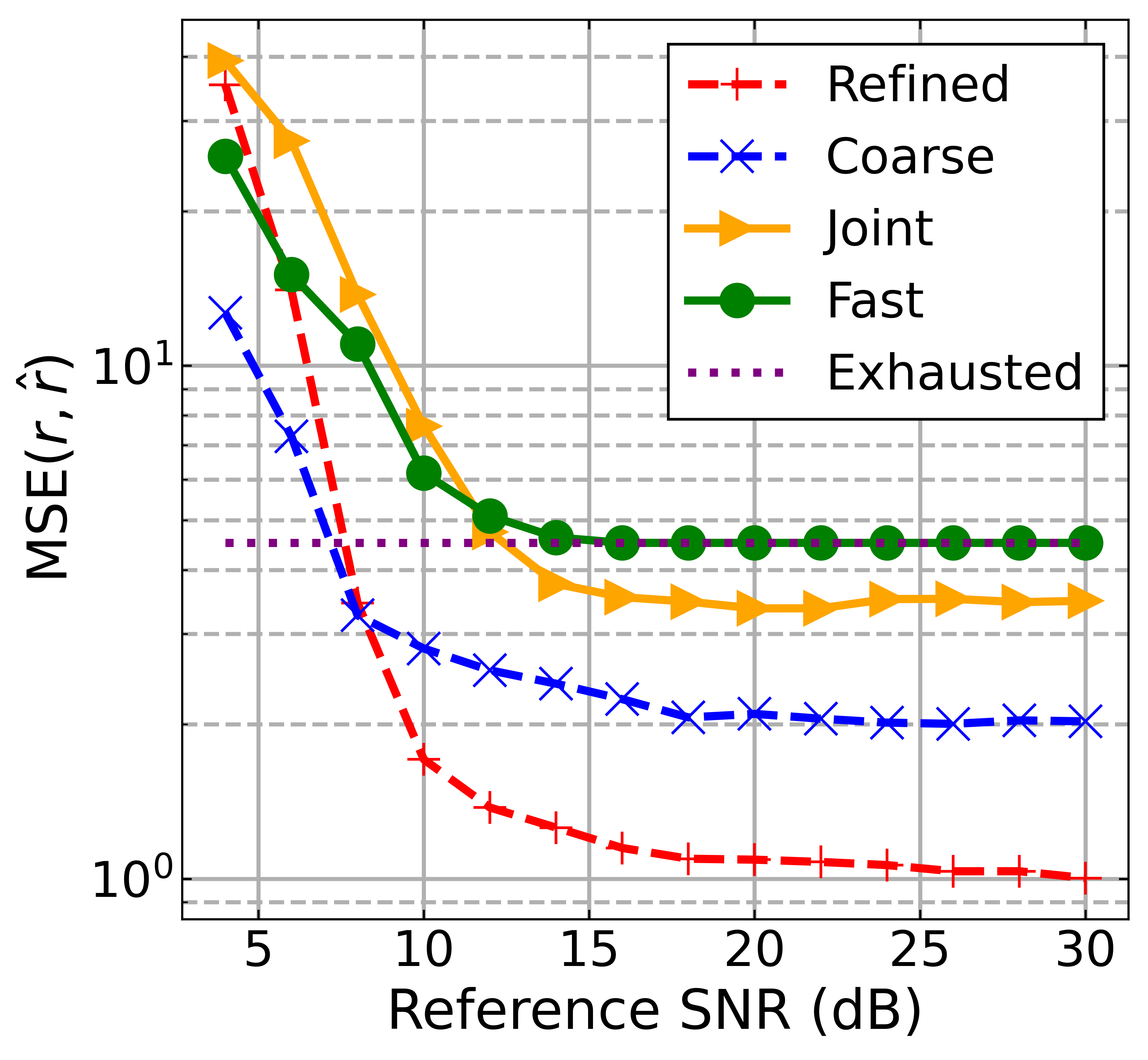}%
    \label{Distance MSE}%
}
\caption{MSE performance of (a) angle and (b) distance estimation versus reference SNR.}
\label{fig:MSEperformance}
\end{figure}

\subsubsection{Achievable Rate}Secondly, we evaluate the achievable rate using different beam training schemes. We begin with the single-user scenario. We also consider full channel information as a baseline, for which the true position $(\theta,r)$ is known and $\mathbf{v}=\mathbf{b}(\theta,r)$ is set as beamforming codeword. The result is shown in \Cref{Achievable rate versus reference SNR}. Overall, the refined method exhibits performance comparable to other schemes, with a gap of up to $0.45 \text{ bits/s/Hz}$ relative to full CSI. Our refined scheme achieves the highest achievable rate in high-SNR scenarios, offering an improvement of up to $0.13\text{ bits/s/Hz}$ compared with the exhausted scheme and $0.04\text{ bits/s/Hz}$ campared with coarse scheme.
\begin{figure}[!t]
\captionsetup{justification=raggedright, singlelinecheck=false} 
\centering
\includegraphics[width=2.5in]{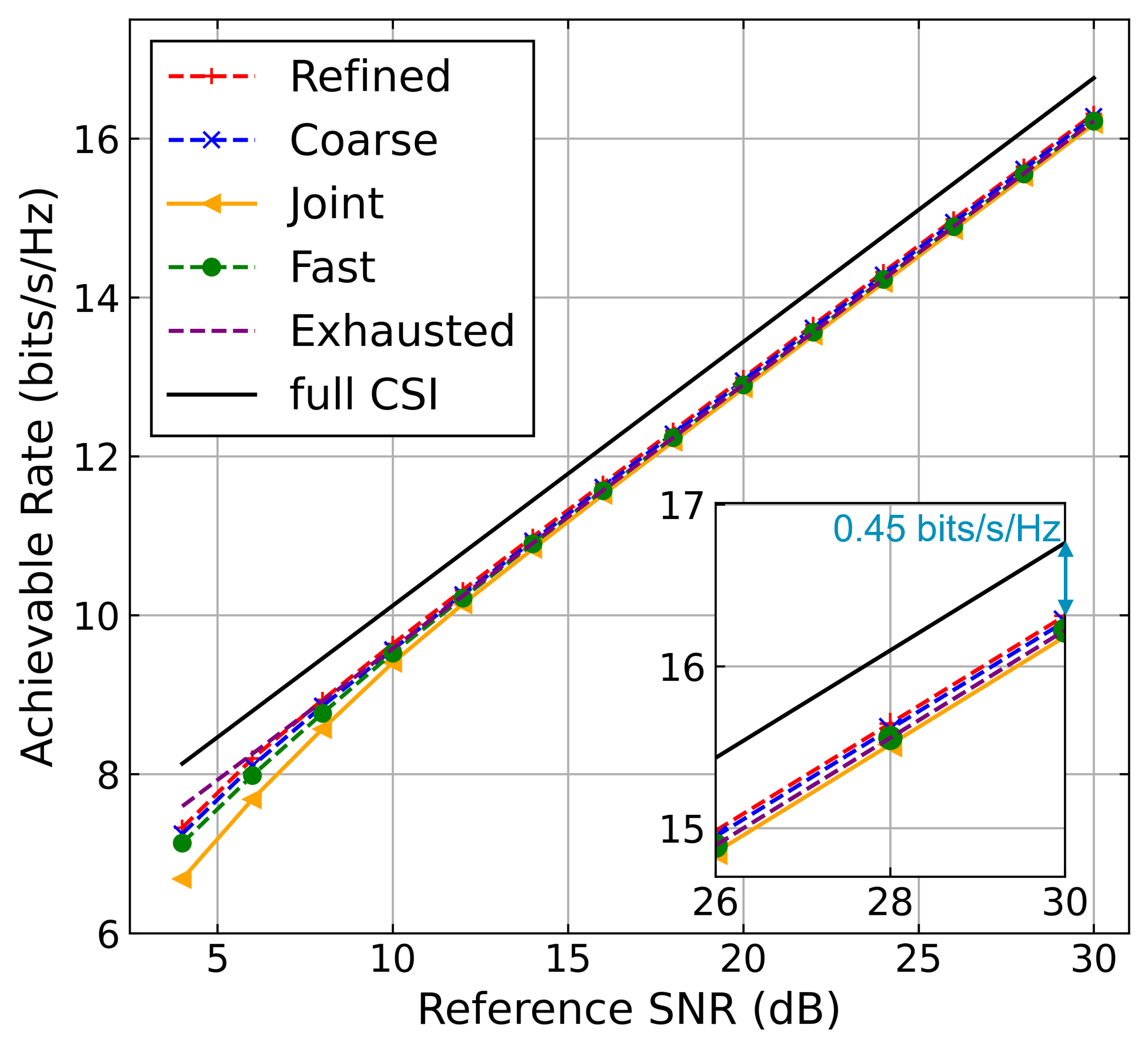}
\caption{Achievable rate versus reference SNR of single-user beamforming}
\label{Achievable rate versus reference SNR}
\end{figure}

In a multi-user scenario, we randomly group \(M = 10\) users within near-field region and perform digital beamforming following the method in \cite{Beam_focusing_multiuser}, thereby serving all \(M\) users within a single time slot. We calculate the average achievable rate for performance evaluation. For each user \(i\) with estimated position \((\hat{\theta}_i, \hat{r}^\rho_i)\) (and corresponding channel \(\mathbf{h}(\theta_i, r_i)\)), the achievable rate is computed based on the SINR as
\begin{equation}
R_i = \log_2\!\Biggl(1+\frac{\bigl|\mathbf{h}^H(\theta_i, r_i)\,\mathbf{v}_i\bigr|^2}{\sum_{j\neq i}\bigl|\mathbf{h}^H(\theta_i, r_i)\,\mathbf{v}_j\bigr|^2 + \sigma^2}\Biggr),
\end{equation}
where \(\mathbf{v}_i\) is the beamforming vector allocated to the user at position \((\hat{\theta}_i, \hat{r}^\rho_i)\) denotes the noise power, and the denominator represents the interference from other users. For full channel information scheme, the true position \((\theta, r)\) is known and thus the beamforming codeword is optimized following the method in \cite{Beam_focusing_multiuser} with true position. The results are shown in \Cref{Achievable rate versus reference SNR2}. The refined method achieves the best performance at high SNR value, with improvements of up to $0.30 \text{ bits/s/Hz}$ over the fast and exhausted schemes, and $0.38$ bits/s/Hz compared to the joint scheme. The refined scheme outperforms the coarse scheme by $0.07 \text{ bits/s/Hz}$. The gap between full CSI and our refined scheme arises from its heavy reliance on precise channel information in multi-user beamforming.
Small errors in angle and distance estimation can lead to significant rate loss, especially when SNR is high. On the other hand, our method is more precise than other schemes as demonstrated in \Cref{fig:MSEperformance}, which leads to the improvement in the achievable rate in this case. For instance, to achieve the rate of $9.25$ bits/s/Hz, our proposed scheme achieves a SNR gain of 2.38 dB compared to exhausted scheme.
\begin{figure}[!t]
\captionsetup{justification=raggedright, singlelinecheck=false} 
\centering
\includegraphics[width=2.5in]{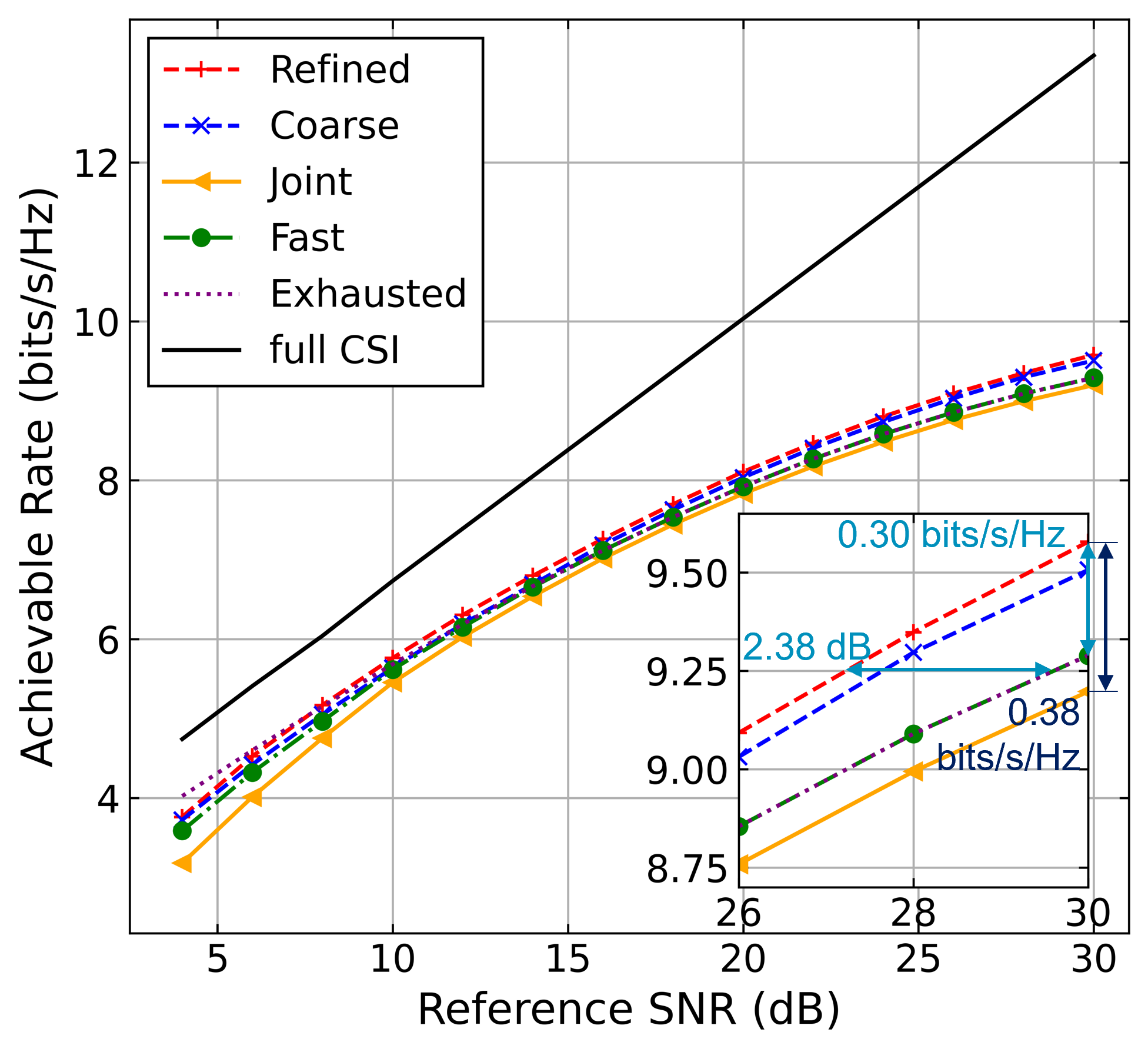}
\caption{Achievable rate versus reference SNR of multi-user beamforming}
\label{Achievable rate versus reference SNR2}
\end{figure}

\subsubsection{Sample Rate}Thirdly, we evaluate the impact of DFT codebook sample rate on different schemes. The sample rate of DFT codebook denotes the number of uniformly sampled angles. As stated in \cite{Joint_Angle_and_Range_Estimation}, the joint scheme is sensitive to the sample rate of the DFT codebook. On the other hand, the performance of joint scheme will increase using the over-sampling strategy. Thus we compare the performance of coarse scheme, refined scheme and joint scheme using different sample rates. We set the reference SNR as $20$ dB to avoid the influence of noise. 
We test the MSE of distance and angle on 100 NUs for joint scheme, proposed scheme and refined scheme. The results are shown in \Cref{table:theta_mse,table:R_mse} It can be seen that although performances of all three methods vary with the sample rate, the coarse and refined method still outperformed the joint scheme.
\begin{table}[!h]
\caption{Theta MSE Values at Different Sample Rates}
\label{table:theta_mse}
\centering
\begin{tabular}{c c c c}
\toprule
Sample Rate & Joint & Coarse & Refined \\
\midrule
512  & $1.4202\times10^{-6}$ & $1.4301\times10^{-6}$ & $1.4301\times10^{-6}$ \\
1024 & $4.8100\times10^{-7}$ & $4.0370\times10^{-7}$ & $4.0590\times10^{-7}$ \\
2048 & $1.1503\times10^{-6}$ & $3.3890\times10^{-7}$ & $3.4000\times10^{-7}$ \\
\bottomrule
\end{tabular}
\end{table}

\begin{table}[!h]
\caption{R MSE Values at Different Sample Rates}
\label{table:R_mse}
\centering
\begin{tabular}{c c c c}
\toprule
Sample Rate & Joint & Coarse & Refined \\
\midrule
512  & 4.1551 & 2.3140 & 1.5453 \\
1024 & 3.9710 & 1.3571 & 0.8578 \\
2048 & 2.1101 & 0.9787 & 0.9996 \\
\bottomrule
\end{tabular}
\end{table}

\begin{figure}[!t]
\captionsetup{justification=raggedright, singlelinecheck=false} 
\centering
\includegraphics[width=2.5in]{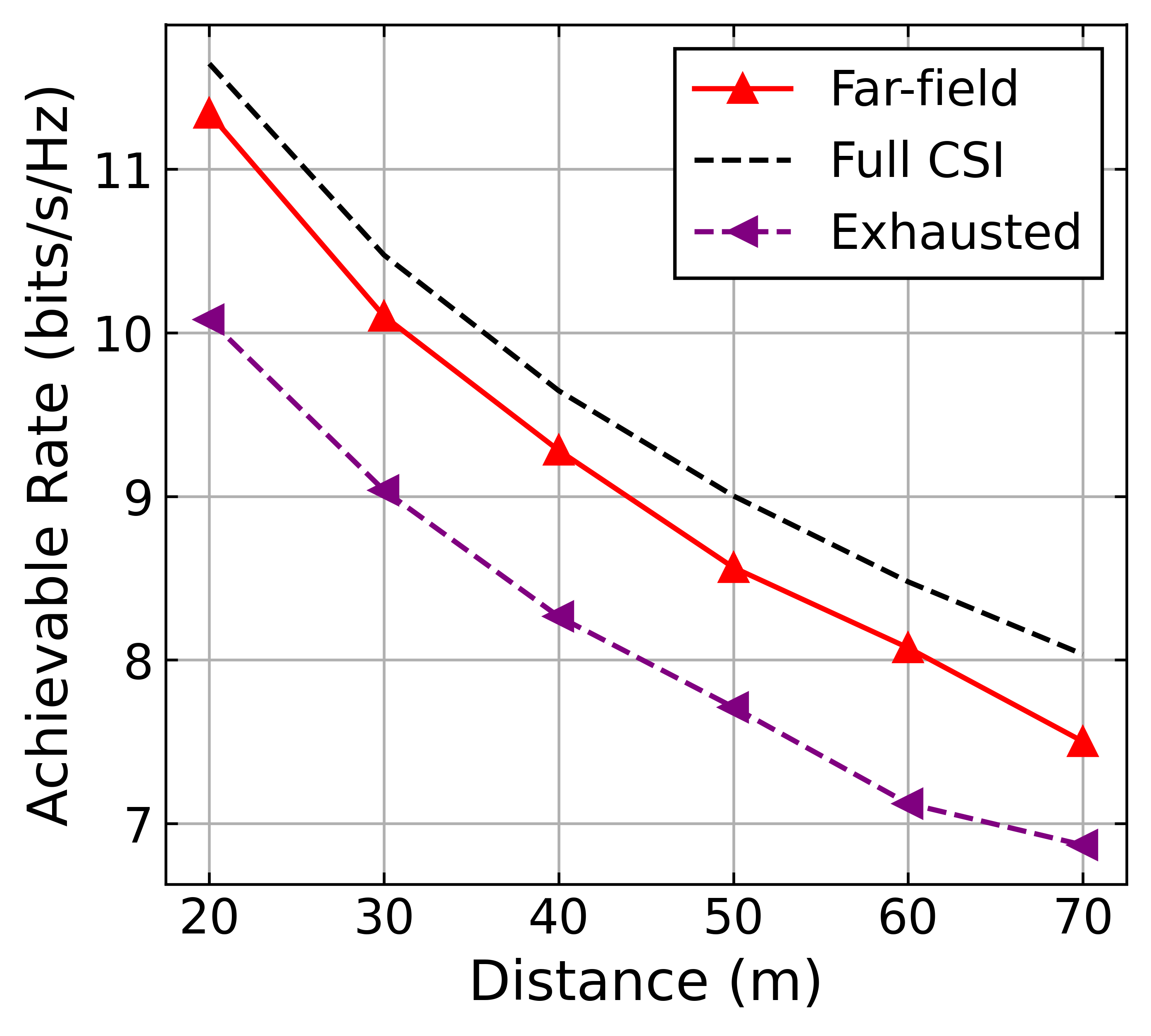}
\caption{Achievable rate of far-field users versus Distance}
\label{fig:Achievable rate of FU}
\end{figure}

\subsubsection{Modified Rayleigh distance}Fourthly, we evaluate the effectiveness of our proposed  
modified Rayleigh distance. We select FUs near the boundaries of the modified Rayleigh distance at different angles $\theta$ (see \Cref{Rayleigh_theoretical}), and then compare the average achievable rate of beamforming with far-field codeword and near-field codeword (full CSI) for 500 single-user cases. Specifically, we analyze whether users classified as FUs under our modified Rayleigh distance achieve a high achievable rate when using only the far-field codeword. The reference SNR is set at $20$ dB to minimize noise effects. The results are shown in \Cref{fig:Achievable rate of FU}. The far-field DFT beamforming vector $\mathbf{a}(\hat{\theta})$ outperforms the near-field codeword $\mathbf{b}(\hat{\theta},\hat{r}^\rho)$ obtained via the exhaustive search, which are respectively labeled as "Far-field" and "Exhausted." 
It is worth noting that the achievable rate of the far-field scheme is lower than the near-field beam forming vector $\mathbf{b}(\theta,r)$ using the user's true location $(\theta,r)$, which is labeled as "Full CSI", with the worst-case performance at approximately $93\%$. Exhausted scheme samples less in the far-field regions and therefore can not achieve a good achievable rate.

\subsubsection{Overhead and Complexity}
Finally, we compare the beam sweeping overhead and complexity. The results can be seen in \Cref{table:overhead}, where $k$ median angles are selected, $S>1$ is the average sampled distance at different angle in the near-field codebook, and  $N$ is the number of angles sampled in the angular domain, which is set the same for DFT and near-field codebook for fair comparison. The overhead of our schemes are the lowest since we only samples in angular domain for beam training. 
Then we compare the complexity of different schemes 
 as shown in \Cref{table:overhead}. The 
 exhausted scheme has one single stage and the other schemes have two stages. 
 Although the joint scheme and proposed scheme have the same sweeping overhead, the proposed schemes have the lowest complexity. 
The complexity of proposed schemes in the angle estimation stage is the same as that of the joint and fast scheme, which is $\mathcal{O}(N)$. 
However, our coarse scheme exhibits a lower complexity in the distance estimation stage. According to \cite{Joint_Angle_and_Range_Estimation}, the complexity of joint scheme is $\mathcal{O}\left(\left|\mathcal{Z}_{\mu}\right| N\right)$, where $\mathcal{Z}_{\mu}>1$ is a factor for the search by distance, while the proposed coarse method doesn't need to search and can directly calculate the distance. Therefore, the complexity is $\mathcal{O}\left(1\right)$. In the refined scheme, we compute the exact threshold and then perform iterative updates. As shown in \Cref{fig:iteration}, the scheme converges rapidly— the distance MSE stabilizes after roughly four iterations. Therefore, we can safely set the maximum iteration number \(I_{\mathrm{max}}\) to a small constant, yielding a computational complexity of $\mathcal{O}(I_{\mathrm{max}})=\mathcal{O}(1)$.

\begin{figure}[!t]
\captionsetup{justification=raggedright, singlelinecheck=false} 
\centering
\includegraphics[width=2.25in]{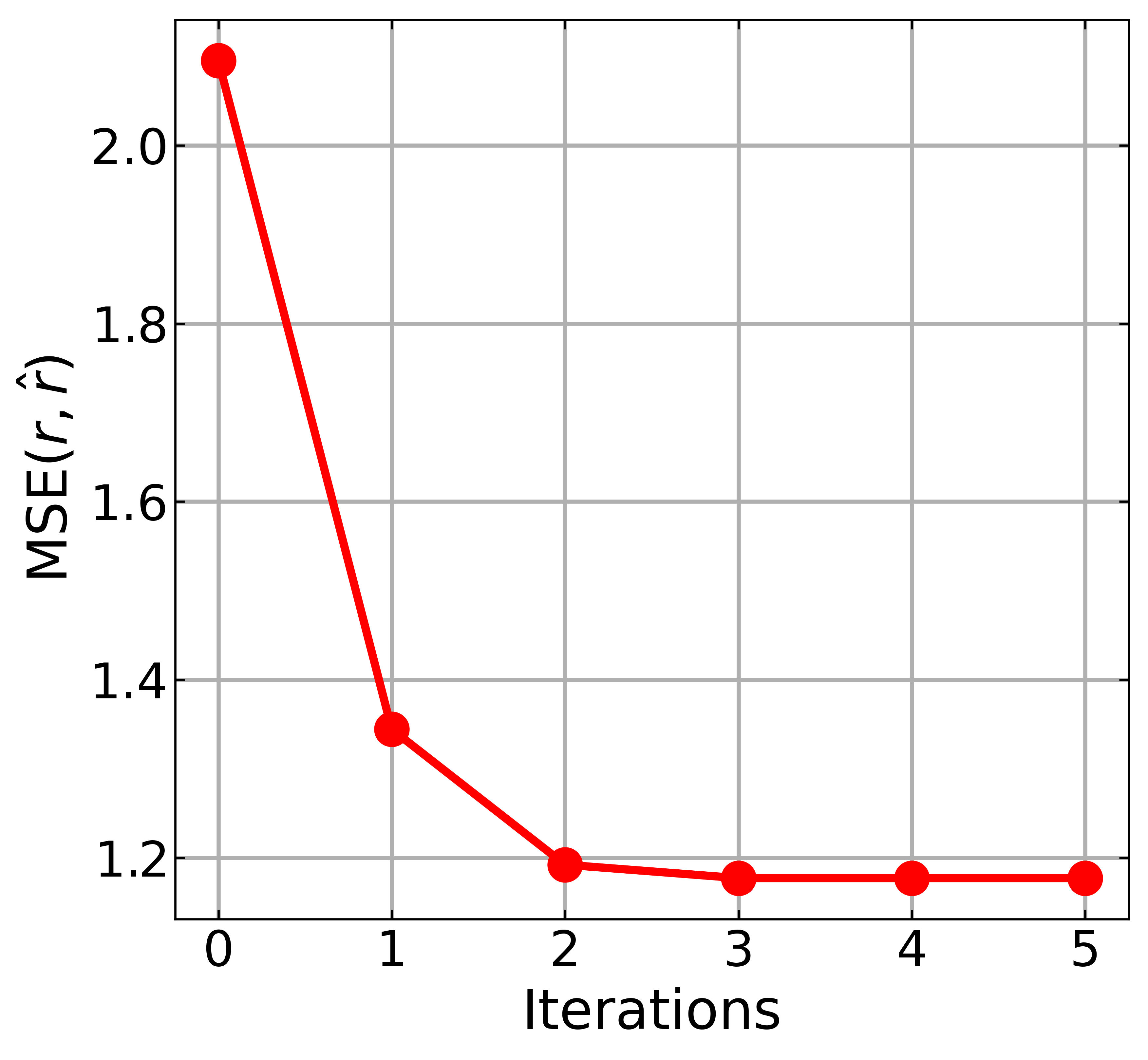}
\caption{MSE of distance in refined scheme versus Iterations}
\label{fig:iteration}
\end{figure}

\vspace{0mm}
\begin{table}[!h]
\caption{Training Overhead and Complexity for Different Beam Training Schemes}
\label{table:overhead}
\centering
\begin{tabular}{l c c}
\toprule
Scheme     & Training Overhead & Complexity\\
\midrule
Refined    & \(N+k\)         & \(\mathcal{O}(N)+\mathcal{O}(I_{\text{max}})\)\\
Coarse     & \(N+k\)           & \(\mathcal{O}(N)+\mathcal{O}(1)\)\\
Joint      & \(N+k\)           & \(\mathcal{O}(N)+\mathcal{O}\left(\left|\mathcal{Z}_{\mu}\right|N\right)\)\\
Fast       & \(N+k\cdot S\)     & \(\mathcal{O}(N)+\mathcal{O}(k\cdot S)\) \\
Exhausted  & \(N\cdot S\)       & \(\mathcal{O}(N\cdot S)\) \\
\bottomrule
\end{tabular}
\end{table}

\section{Maximum Likelihood Estimation Scheme}\label{sec:Proposed Maximum Likelihood Scheme}
To further improve estimation accuracy beyond deterministic low-complexity schemes, we adopt a MLE approach that accounts for the statistical fluctuations of received signal amplitudes. This approach chooses initials based on the beam cluster $\mathcal{C}^*$ defined in \Cref{sec:coarse and refine} and a protect region. We derive the corresponding log-likelihood function and formulate an optimization problem to jointly estimate user location and channel parameters.

\subsection{Problem Formulation}\label{subsec:likelihood}
We denote $Z_n=|y(\mathbf{v_n})|$.
From \Cref{eq:received signal model}, the received signal can be modeled as
\begin{equation}
y(\mathbf{v_n}) = u_n e^{\jmath\psi_n} + w,
\end{equation}
with
\begin{subequations}
\label{eq:signal_params}
\begin{align}
u_n &= |\mathbf{h}^{H}(\theta, r)\mathbf{v}_n x|, \label{eq:signal_params_a}\\[1mm]
\psi_n &= \operatorname{Arg}[\mathbf{h}^{H}(\theta, r)\mathbf{v}_n x]. \label{eq:signal_params_b}
\end{align}
\end{subequations}
 Since the noise \(w\sim \mathcal{CN}(0,\sigma^2)\), one can have $y(\mathbf{v}_n) \sim \mathcal{CN}(u_n e^{\jmath\psi_n},\sigma^2)$. The distribution of the amplitude of nonzero mean complex gaussian variable is Rice distribution, thus the distribution of $Z_n$ is
\begin{equation}
f_{Z_n}(z_n) = \frac{2z_n}{\sigma^2}\exp\!\left(-\frac{z_n^2+u_n^2}{\sigma^2}\right)I_0\!\left(\frac{2z_nu_n}{\sigma^2}\right),\quad z_n\ge0.
\label{Rice Distribution of Received Signal Strength}
\end{equation}
The detailed proof can be found in \Cref{app:Rice}. Since they are independent with each other. The joint PDF of $Z_n,n\in\mathcal{N}$ is
\begin{equation}
f_{Z_0,Z_1,\cdots,Z_{N-1}}(z_0,z_1,\cdots,z_{N-1}) = \prod_{n=0}^{N-1} f_{Z_n}(z_n).
    \label{eq:joint pdf of Z_n}
\end{equation}
From \Cref{eq:channel model,eq:signal_params_a}, and by setting pilot symbol $x=1$, we have the normalized channel gain
\begin{equation}
u_n=\frac{D}{r}\lvert\mathbf{b}^H(\theta,r)\mathbf{a}(\varphi_n)\rvert,
\end{equation}
where $D$ is a real-valued amplitude factor related to channel gain. We consider this factor as a parameter to be estimated by MLE. Let $\mathbf{Z}=(Z_0,Z_1,\cdots,Z_{N-1})$, $\mathbf{\Theta}=(\theta,r,D,\sigma^2)$ the log-likelihood function is
\begin{align}
    \ell(\mathbf{Z};\mathbf{\Theta} ) &= \sum_{n=0}^{N-1} \ln(f_{Z_n}(z_n)) \notag \\
    &= \sum_{n=0}^{N-1} \Biggl\{ \ln(2z_n) - \ln(\sigma^2) - \frac{z_n^2 + u_n^2}{\sigma^2} \notag \\
    &\quad\; + \ln\Bigl(I_0\Bigl(\frac{2z_nu_n}{\sigma^2}\Bigr)\Bigr) \Biggr\}.
    \label{eq:loglikehood}
\end{align}

The optimization problem can be formulated as
\begin{align}
    \min_{\mathbf{\Theta}} \quad & -\ell(\mathbf{Z};\mathbf{\Theta}), \notag \\
    \text{subject to} \quad & \theta \in [-1, 1], \quad r \in [R_{\text{Fre}}, R_{\text{Ray}}],\quad\sigma^2\neq0.
    \label{eq:optimization_problem}
\end{align}
The boundary of $\sigma^2$ and $D$ is relaxed because of the symmetric of $\ell(\mathbf{Z};\mathbf{\Theta})$. \Cref{fig:2D costfunction} and \Cref{fig:1D costfunction} illustrate an example of the log-likelihood function when user is located at $(0,5\text{m})$ and $\text{SNR} = 10$ dB.
\begin{figure}[!t]
\captionsetup{justification=raggedright, singlelinecheck=false} 
\centering
\includegraphics[width=2.5in]{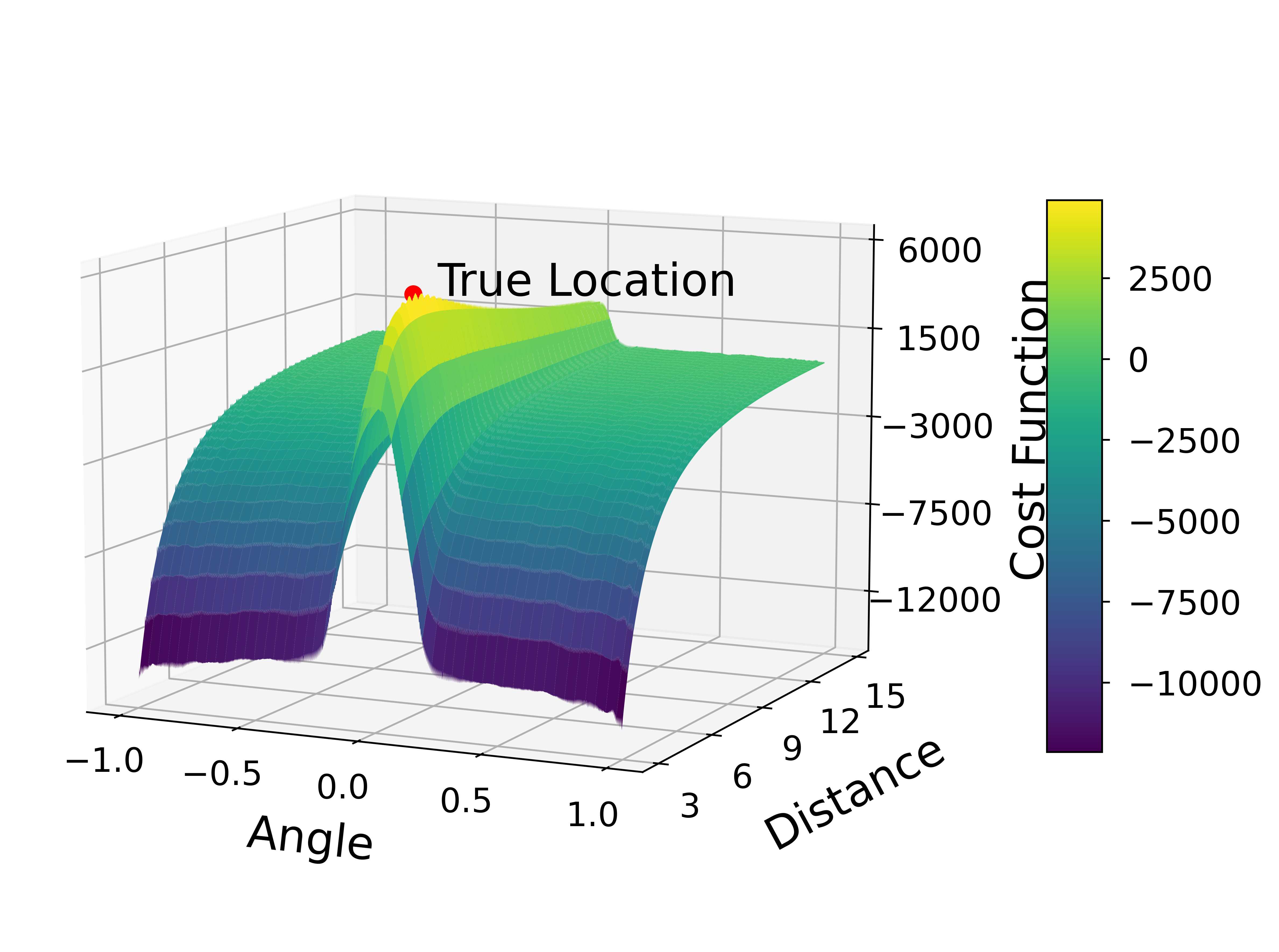}
\caption{The 2D shape of log-likelihood function with respect to angle and distance for user at (0,5\text{m}).}
\label{fig:2D costfunction}
\end{figure}
\begin{figure}[!t]
\captionsetup{justification=raggedright, singlelinecheck=false} 
\centering
\includegraphics[width=2.5in]{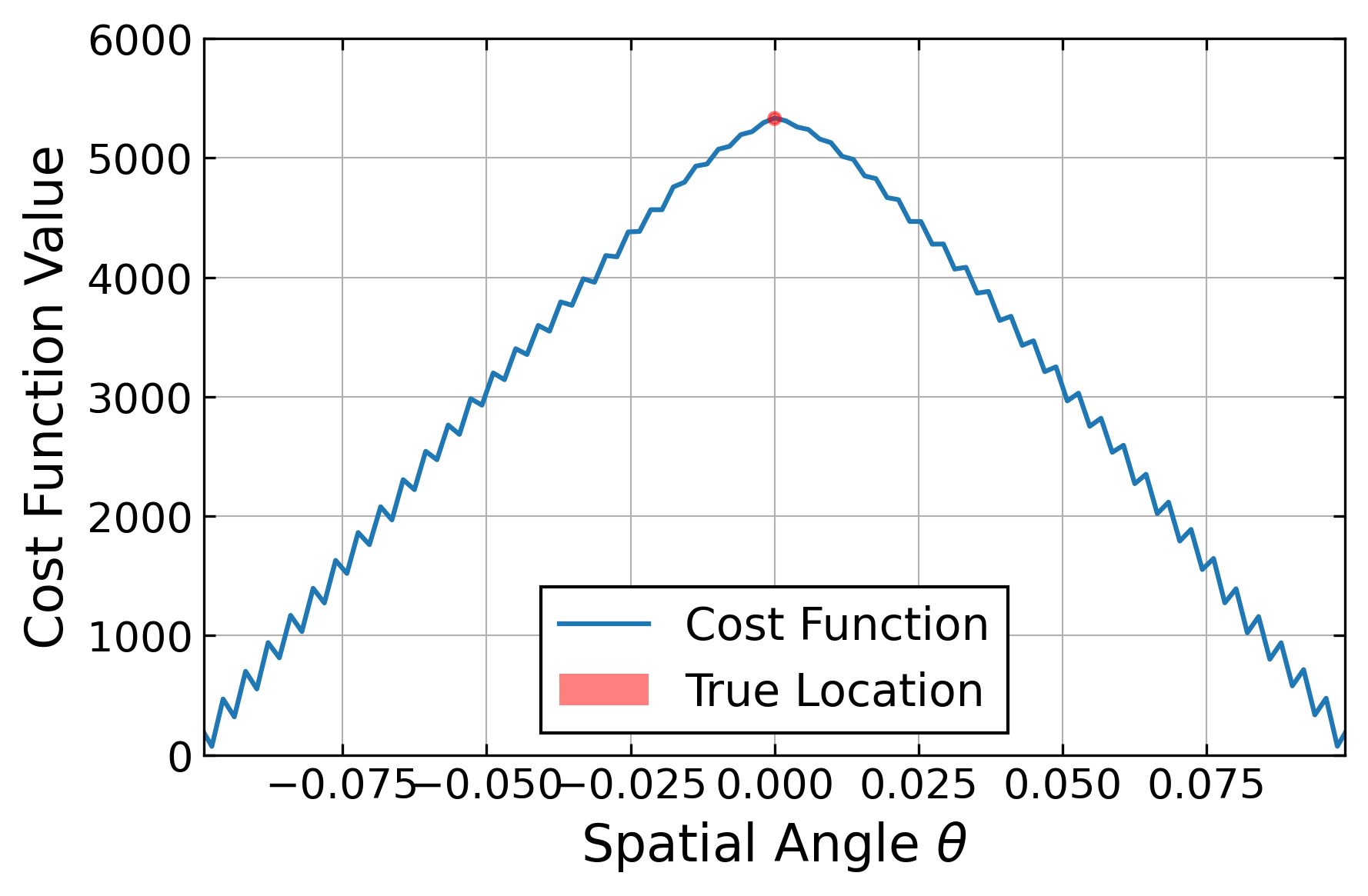}
\caption{The 1D shape of log-likelihood function with respect to angle at 5m for user at (0,5\text{m}).}
\label{fig:1D costfunction}
\end{figure}

As the diagrams show, the log-likelihood function involves some small-scale fluctuations, and the vanilla gradient descent (GD) may get stuck. Instead, we combine the Adam method in paper \cite{AdamAM_GD} and the SA method in \cite{bertsimas1993simulated} to search for the optimal solution. The SA stage helps avoid poor local optima by introducing stochastic exploration via a decaying temperature schedule, while the Adam stage exploits gradient statistics to efficiently refine the estimate.

\begin{figure}[!t]
\captionsetup{justification=raggedright, singlelinecheck=false} 
\centering
\includegraphics[width=2.5in]{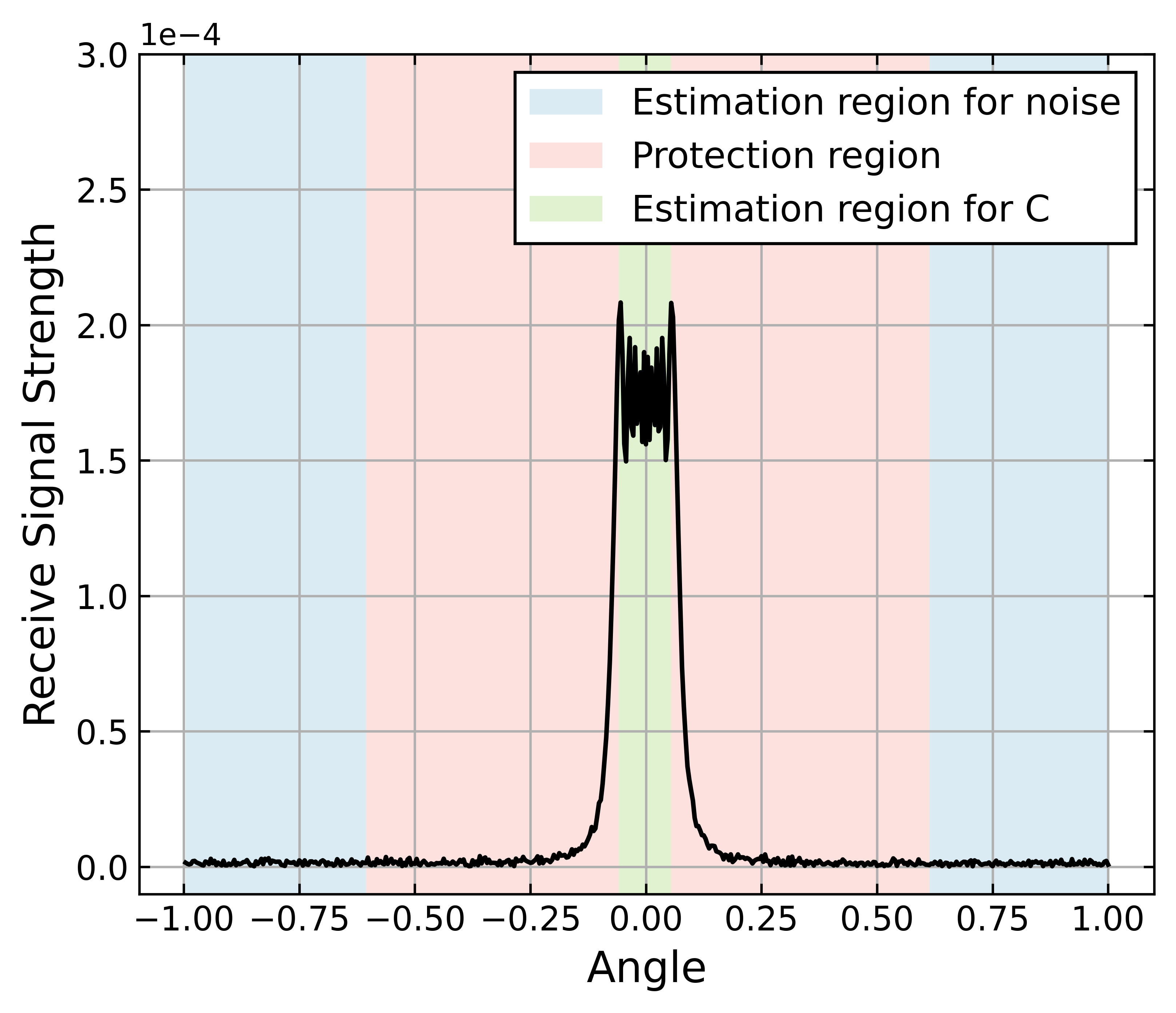}
\caption{The region of selected beam pattern for different estimation for user at $(0,5\text{m})$.}
\label{fig:region}
\end{figure}
\subsection{Initial estimation of the Amplitude Factor \(D\) and noise power $\sigma^2$}\label{subsec:pre-est}
We obtain the initial values of $\hat{\theta}_0$ and $\hat{r}_0$ from the coarse scheme. Then we obtain initial values $\hat{D}_0$ as follows. 

The deviation of the log-likelihood function w.r.t $D$ is represented as
\begin{equation}
\frac{\partial \ell(\mathbf{Z};\mathbf{\Theta} )}{\partial D} = \sum_{n=0}^{N-1} \left[-\frac{D\,A_n^2}{\sigma^2} + \frac{2z_n\,A_n}{\sigma^2}\,\frac{I_1\!\left(\frac{2z_n D\,A_n}{\sigma^2}\right)}{I_0\!\left(\frac{2z_n D\,A_n}{\sigma^2}\right)}\right] = 0,
\label{eq:deriv_of_C}
\end{equation}
where $A_n=u_n/D$.
When $o$ is large, we can employ the approximation
\begin{equation}
\frac{I_1(o)}{I_0(o)} \approx 1 - \frac{1}{2o},\quad o=\frac{2z_n D\,A_n}{\sigma^2},
\end{equation}
To ensure that the approximation \( \frac{I_1(o)}{I_0(o)} \approx 1 - \frac{1}{2o} \) holds, we require a sufficiently large value of \( o = \frac{2z_n D\,A_n}{\sigma^2} \), which is proportion to SNR in high SNR region. Hence, for estimating \(D_0\), we only select the beams with the strongest received powers to guarantee high SNR. Conveniently, we reuse the beam cluster \( \mathcal{C}^* \) already identified in the coarse scheme. Let \( \mathcal{H} \subset \mathcal{C}^* \) be a subset of \(\mathcal{C}^*\) obtained by removing a few edge elements (protection region) on both sides to form a reliable estimation region and avoid boundary effects. The estimation of \(D_0\) is then based on \( \mathcal{H} \) by 
\begin{equation}
\hat{D}_0 \;=\; \frac{\displaystyle \sum_{n \in \mathcal{H}} z_n\,A_n \;+\; \sqrt{\left(\sum_{n \in \mathcal{H}} z_n\,A_n\right)^2 - |\mathcal{H}|\,\sigma^2\,\sum_{n \in \mathcal{H}} A_n^2}}{2\,\sum_{n \in \mathcal{H}} A_n^2},
\end{equation}
where \( |\mathcal{H}| \) denotes the number of selected beams. The selection region is illustrated in \Cref{fig:region}, where the green portion indicates the index set \( \mathcal{H} \). The resulting MSE performance of this pre-estimation is shown in \Cref{fig:C} (purple line), with an estimation error around \(10^{-5}\) across all SNRs.

On the other hand, to estimate the noise power \( \sigma^2 \), we utilize the beams outside the cluster \( \mathcal{C}^* \), excluding a few adjacent beams (protection region) to avoid contamination from useful signals. Let \( \mathcal{L} \subset \mathcal{N} \setminus \mathcal{C}^* \) be the set of selected indices for this purpose. The initial estimate is computed by 
\begin{equation}
\hat{\sigma_0^2} = \frac{1}{|\mathcal{L}|} \sum_{n \in \mathcal{L}} z_n.
\end{equation}
As shown in \Cref{fig:region} (blue area), this selection ensures low signal components and yields a highly accurate estimate. The MSE of \( \hat{\sigma_0^2} \) is presented in \Cref{fig:sigma2}, achieving error levels below \(10^{-19}\) across SNRs.

\subsection{Derivation of Estimation Error Lower bound}
In this section, we derive CRB for MLE scheme, which is an lower bound for performance evaluations. We first denote the score function with the gradient of the log-likelihood function
\begin{equation}
\nabla_{\mathbf{\Theta}} \ell(\mathbf{Z};\mathbf{\Theta})
= \bigl[\,
\tfrac{\partial \ell(\mathbf{Z};\mathbf{\Theta})}{\partial \theta},\;
\tfrac{\partial\ell(\mathbf{Z};\mathbf{\Theta})}{\partial r},\;
\tfrac{\partial \ell(\mathbf{Z};\mathbf{\Theta})}{\partial D},\;
\tfrac{\partial \ell(\mathbf{Z};\mathbf{\Theta})}{\partial \sigma^2}
\bigr]^T.
\end{equation}

The Fisher information matrix \(I(\mathbf{\Theta})\) is computed by using Outer Product of the Score.  
   \begin{equation}
   I(\mathbf{\Theta})
   = \mathbb{E}\Bigl[\,
     \bigl(\nabla_{\mathbf{\Theta}} \ell(\mathbf{Z};\mathbf{\Theta})\bigr)
     \bigl(\nabla_{\mathbf{\Theta}} \ell(\mathbf{Z};\mathbf{\Theta})\bigr)^T
   \Bigr].
   \end{equation}
Because \(\{Z_n\}\) are independent given \(\mathbf{\Theta}\), the total Fisher information is the sum of each sample’s contribution:
\begin{equation}
I(\mathbf{\Theta})
= \sum_{n=0}^{N-1} I_n(\mathbf{\Theta}).
\end{equation}
Thus the dimension of derivation is downsized. The  CRB for unbiased estimators states that the covariance matrix of any unbiased estimator \(\hat{\mathbf{\Theta}}\) cannot be smaller than the inverse of the Fisher information:
\begin{equation}
\operatorname{Cov}\bigl(\hat{\mathbf{\Theta}}\bigr)
\succeq
I(\mathbf{\Theta})^{-1}.
\end{equation}
This serves as the MSE lower bound for MLE. 
\subsection{Performance Evaluation}\label{subsec:MLE}
We will test the performance of proposed MLE scheme and analyze the CRB in this section. The simulation was performed with 10 random NUs, each subjected to 100 independent noise realizations. For each user in each run, only a single-shot received signal ($N=512$) is considered—just as in the low-complexity scheme. Finally, the performance is computed by averaging over all runs and all users.
\subsubsection{MSE Performance}
Using MSE as metrics, \Cref{fig:group} presents a benchmark of the estimation accuracy among the proposed scheme, the coarse scheme, and CRB. It can be seen that MLE scheme achieves much lower MSE in comparison with the coarse scheme and its performance is approaching CRB.
\begin{figure}[!t]
  \centering
  \subfloat[\tiny(a)][\textrm{\small MSE for angle}]{%
    \includegraphics[width=0.47\linewidth]{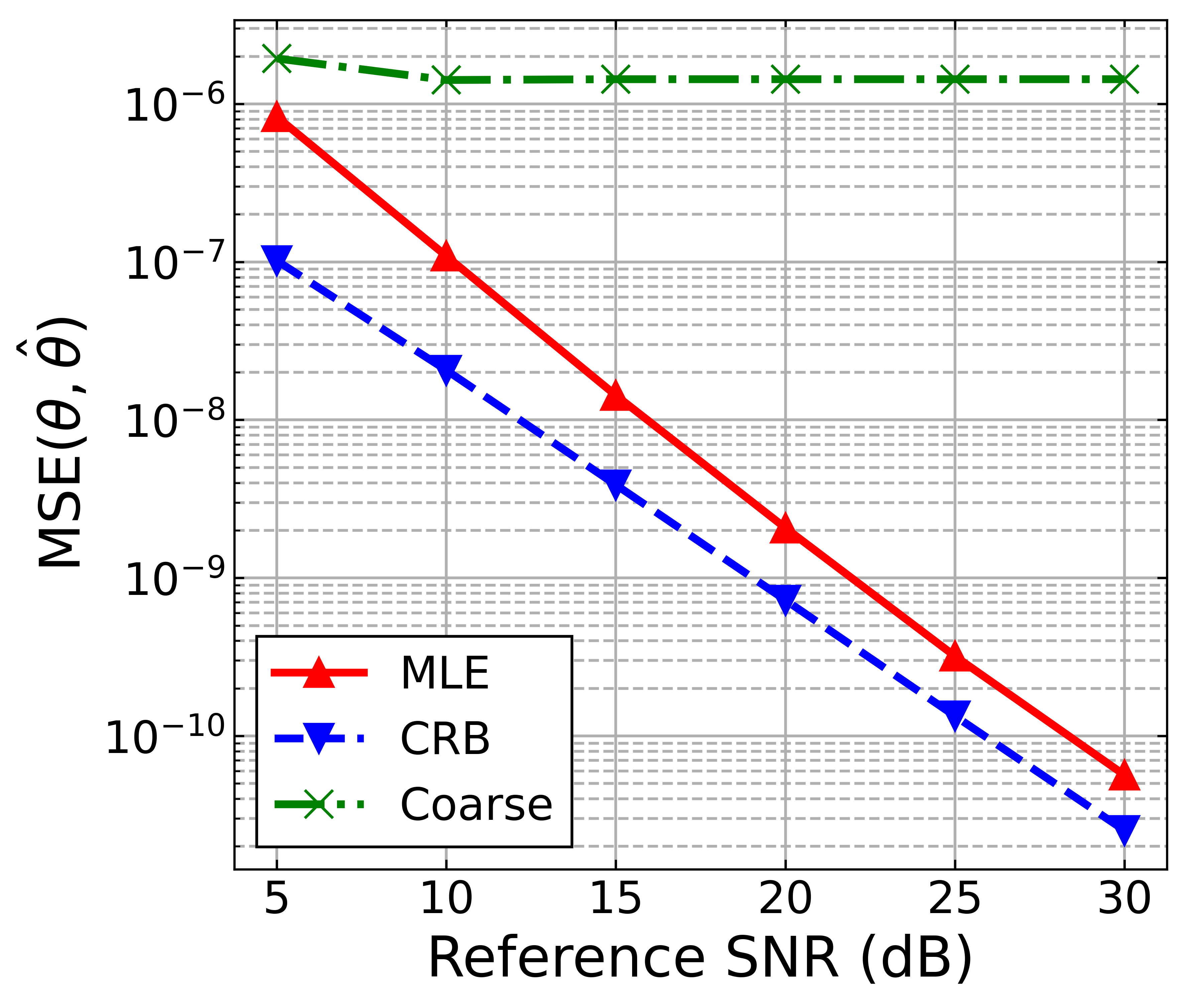}%
    \label{fig:angle}}%
  \quad
  \subfloat[\tiny(b)][\textrm{\small MSE for distance}]{%
    \includegraphics[width=0.47\linewidth]{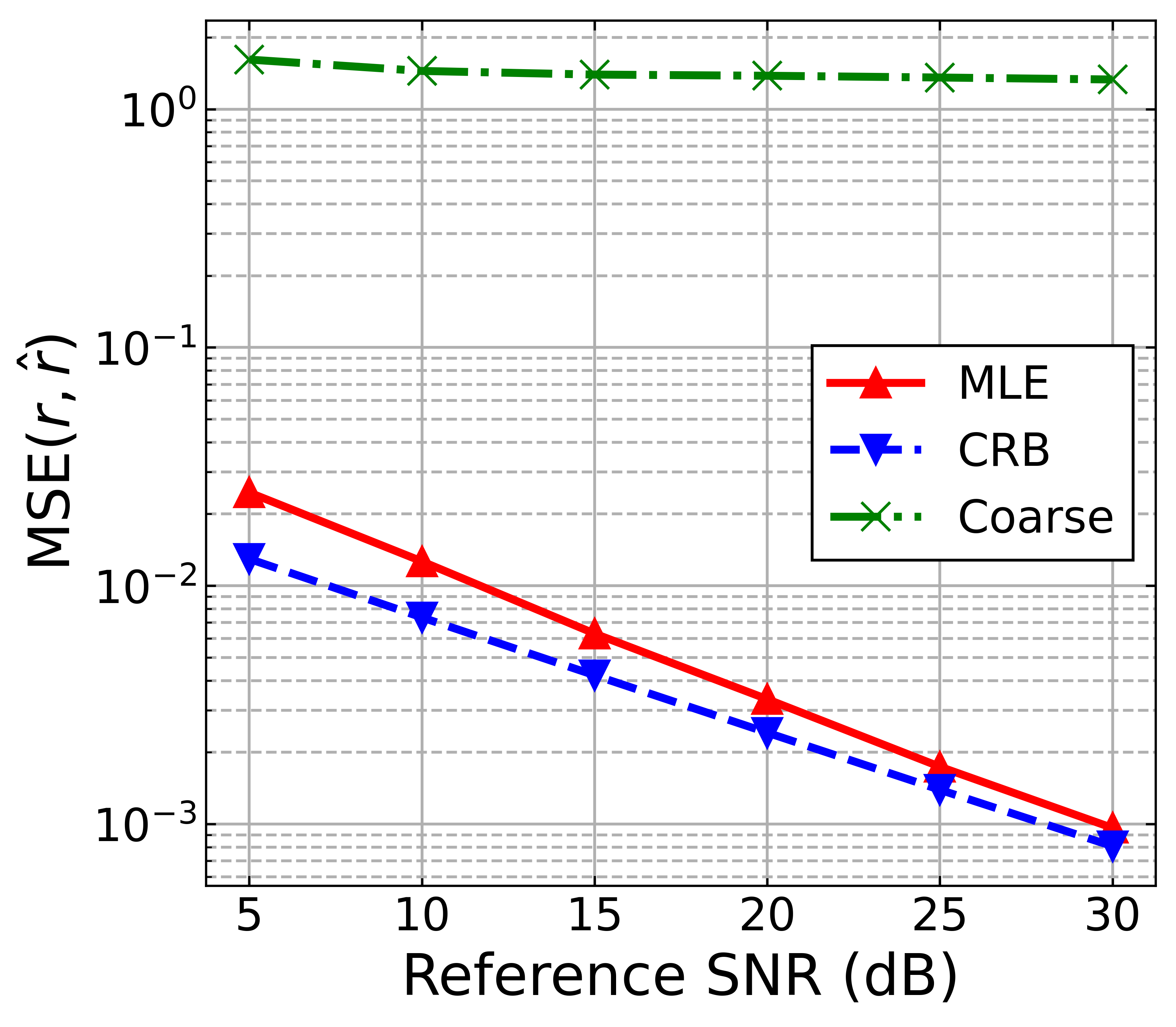}%
    \label{fig:diatance}}%
  
  
  \subfloat[\tiny(c)][\textrm{\small MSE for amplitude factor}]{%
    \includegraphics[width=0.47\linewidth]{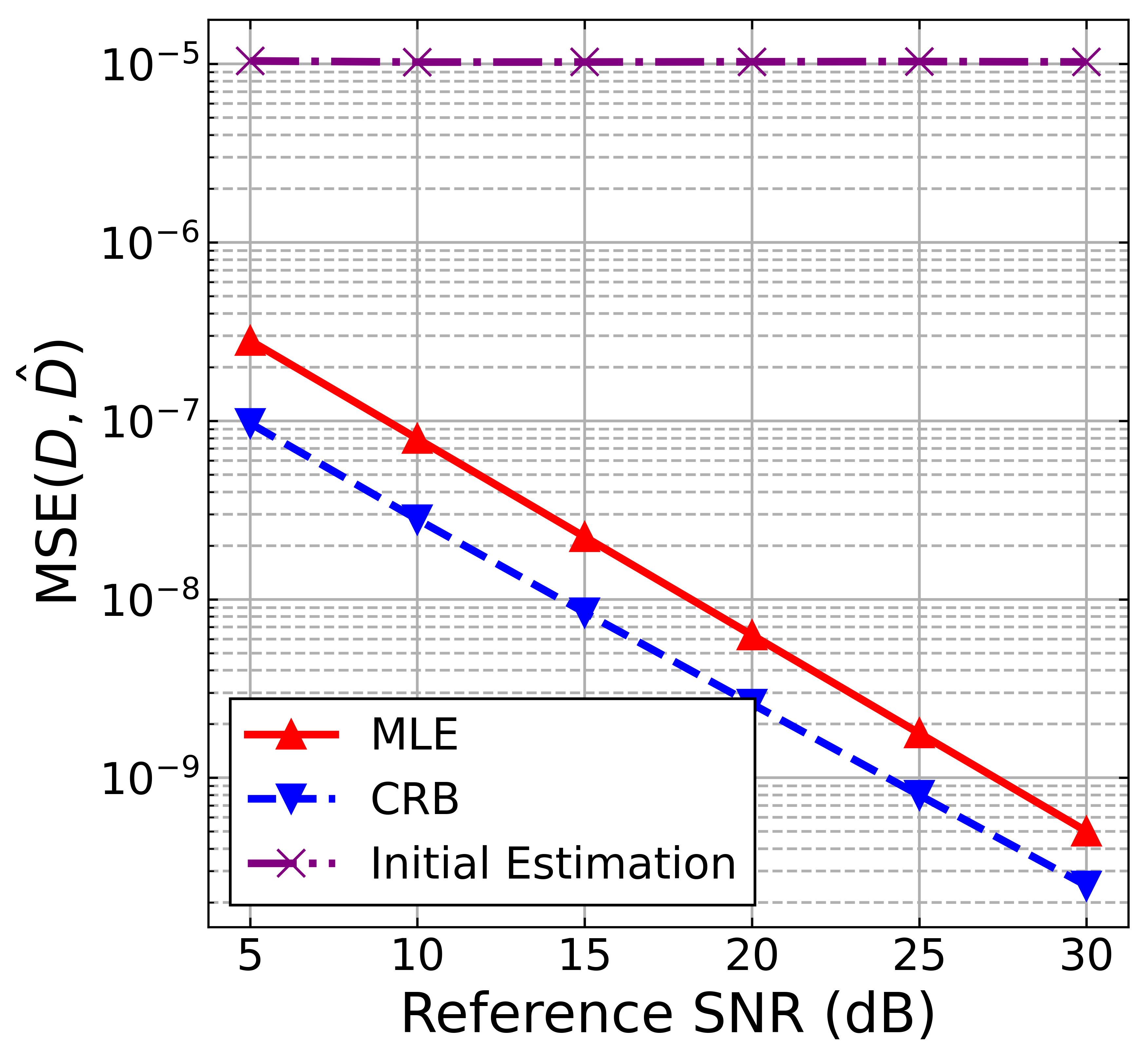}%
    \label{fig:C}}%
  \quad
  \subfloat[\tiny(d)][\textrm{\small MSE for noise power}]{%
    \includegraphics[width=0.47\linewidth]{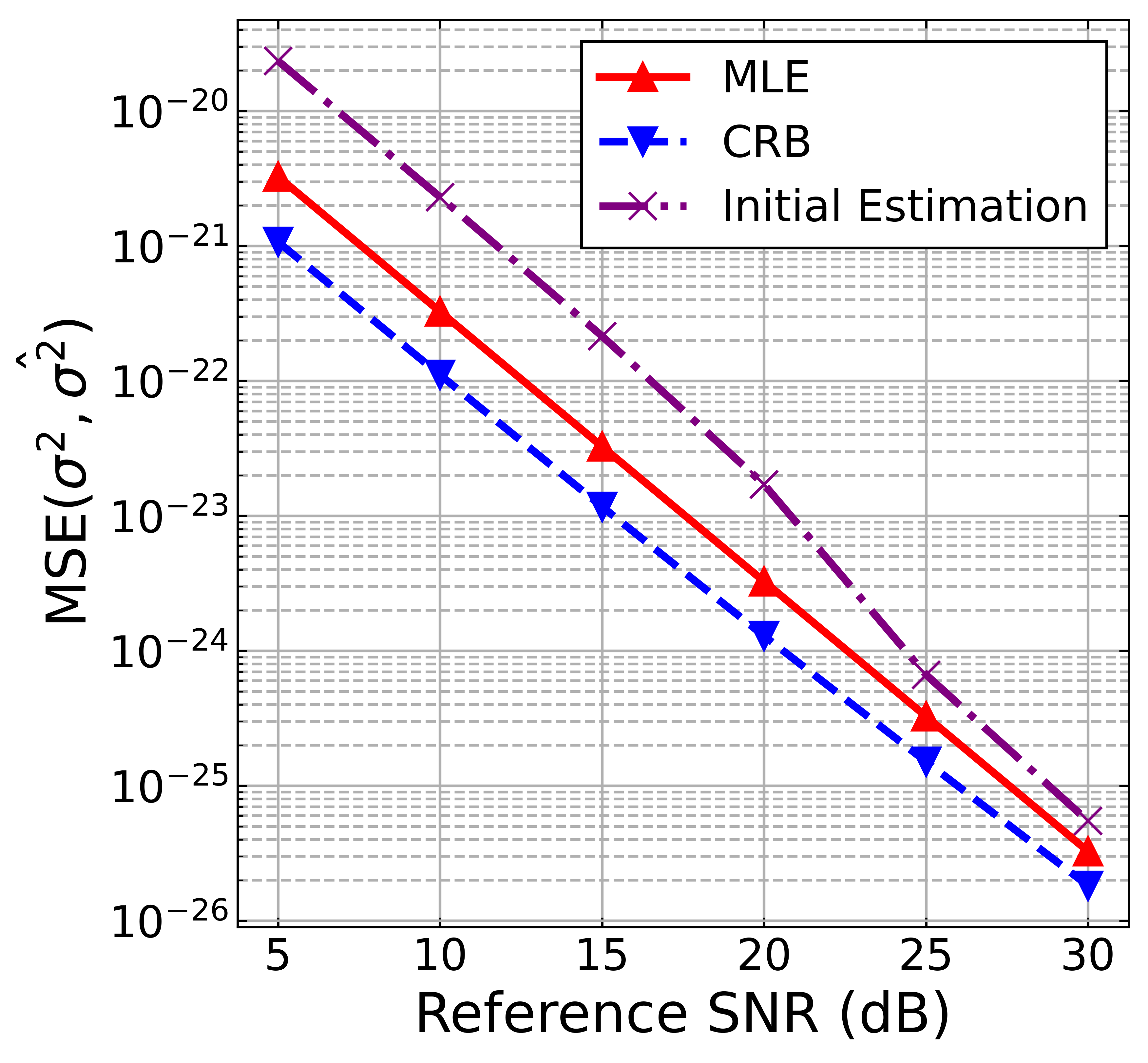}%
    \label{fig:sigma2}}%
  \captionsetup{justification=raggedright, singlelinecheck=false} 
  \caption{The comparison of MSE results of MLE.}
  \label{fig:group}
\end{figure}

\subsubsection{Achievable Rate Performance}
Next, we test the achievable rate via beamforming vector from the refined user locations. After we obtain the user location $(\hat{\theta},\hat{r})$ in $\mathbf{\Theta}^*$, we use near-field codeword $\mathbf{v}=\mathbf{b}(\hat{\theta},\hat{r})$ to perform single-user beamforming. The results are in \Cref{fig:rate}. The MLE scheme has nearly the same achievable rate as the full CSI baseline under all SNRs.
\begin{figure}[!t]
  \centering
  \subfloat[\tiny(a)][\textrm{\small The achievable rate for single-user beamforming.}]{%
    \includegraphics[width=0.45\linewidth]{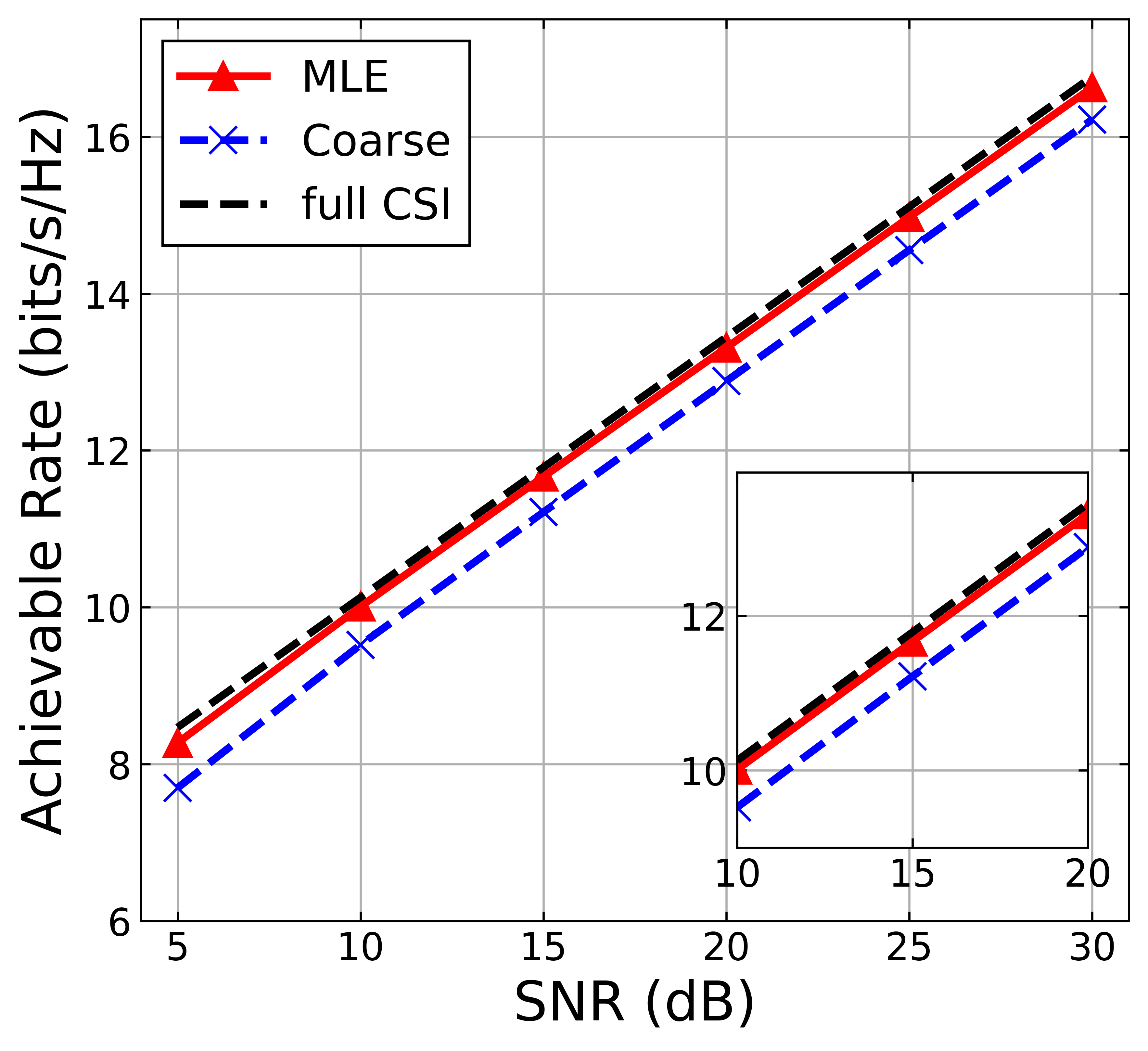}%
    \label{fig:rate}}%
  \quad
  \subfloat[\tiny(b)][\textrm{\small The achievable rate for multi-user beamforming.}]{%
    \includegraphics[width=0.45\linewidth]{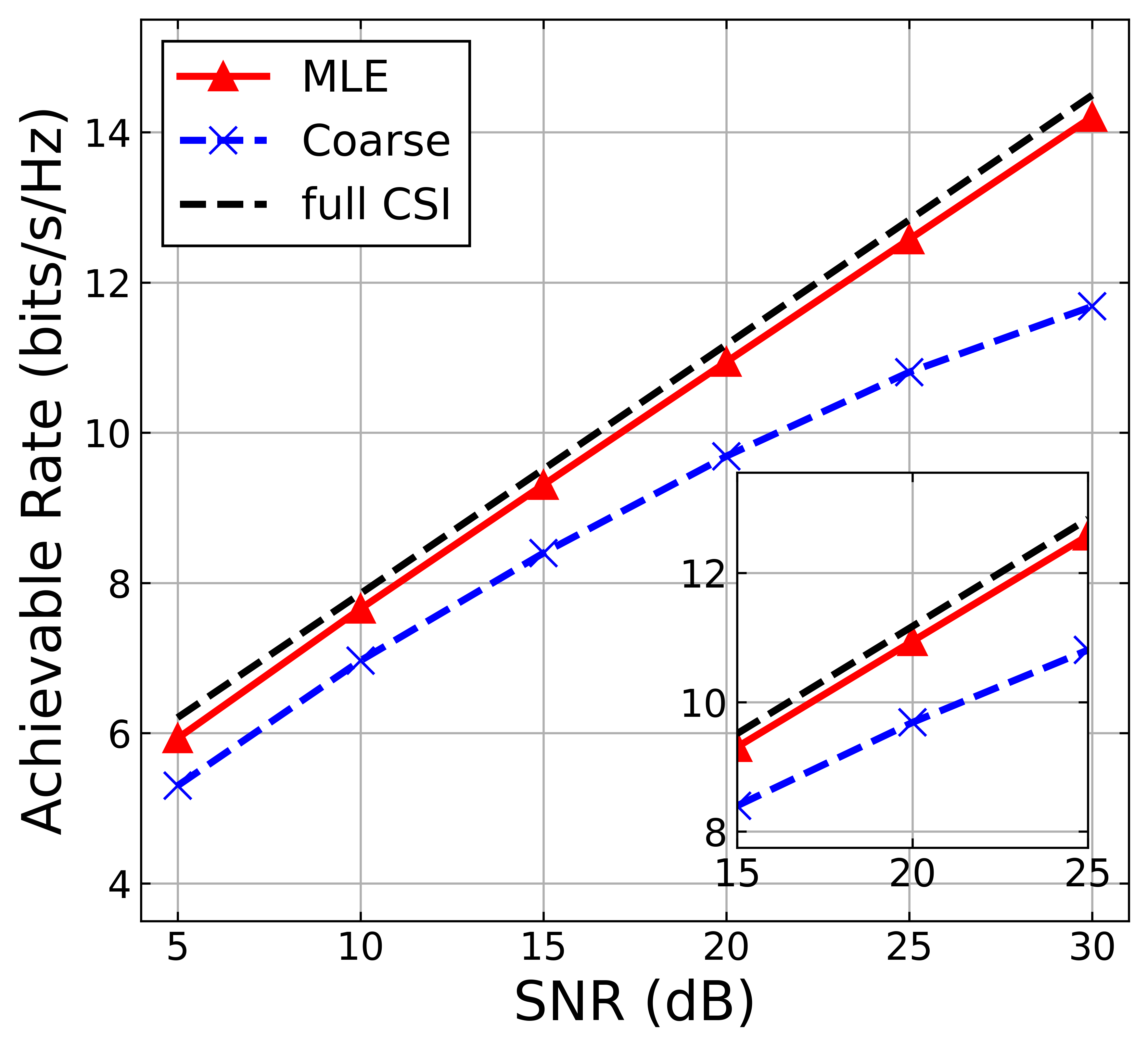}%
    \label{fig:multi_rate}}%
  \captionsetup{justification=raggedright, singlelinecheck=false} 
  \caption{The comparison of achievable rate for different schemes.}
  \label{fig:achievable_rate}
\end{figure}

Then we evaluate the performance of multi-user beamforming for these schemes. We randomly group $M=5$ users and perform beamforming for them simultaneously. \Cref{fig:multi_rate} illustrates that the proposed scheme not only shows significant improvement in the achievable rate but also nearly aligns with the performance of the full CSI scheme in multiple-user scenarios. 
\section{Conclusion}\label{sec:clu}
In this paper, we presented an efficient near-field beam training framework using only far-field DFT codebooks. {By deriving closed-form expressions for beamwidth and central gain, we introduced a modified Rayleigh distance that accurately delineates near- and far-field regimes and underpins our novel two-stage training scheme. Our approach provides precise angle and distance estimates with low complexity, and its performance can be further elevated through an amplitude-only MLE refinement step to approach CRB. Simulation results confirm that the proposed frameworks delivers high estimation accuracy and near-optimal achievable rates in both single- and multi-user scenarios.}

\appendices
\crefalias{section}{appendix}
\section{Derivation of the Closed-Form Expression for \(\tilde{f}(\theta,r,\varphi)\)}\label{app:close-form}

We start from
\begin{equation}
I=\frac{1}{2}\int_{-1}^{1}\exp\Bigl(\jmath\pi\bigl(\alpha x^2-\beta x\bigr)\Bigr)dx,
\end{equation}
Complete square, We have
\begin{equation}
I=\frac{1}{2}\,e^{-\jmath\pi\frac{\beta^2}{4\alpha}}\int_{-1}^{1}\exp\Bigl(\jmath\pi\alpha\Bigl(x-\frac{\beta}{2\alpha}\Bigr)^2\Bigr)dx.
\end{equation}
Let 
\begin{equation}
v=\sqrt{\jmath\pi\alpha}\Bigl(x-\frac{\beta}{2\alpha}\Bigr),
\end{equation}
The integration limits become
\begin{equation}
v_1=\sqrt{\jmath\pi\alpha}\left(-1-\frac{\beta}{2\alpha}\right),\quad
v_2=\sqrt{\jmath\pi\alpha}\left(1-\frac{\beta}{2\alpha}\right).
\end{equation}
Thus,
\begin{equation}
I=\frac{1}{2}\,e^{-\jmath\pi\frac{\beta^2}{4\alpha}}\frac{1}{\sqrt{\jmath\pi\alpha}}
\int_{v_1}^{v_2}\exp(v^2)dv.
\end{equation}
Using the known formula
\begin{equation}
\int \exp(v^2)dv = \frac{\sqrt{\pi}}{2}\operatorname{erfi}(v),
\end{equation}
with \(\operatorname{erfi}(v)=-\jmath\,\operatorname{erf}(\jmath v)\), we obtain the final result
\begin{equation}
\begin{split}
&I=\\
&\frac{e^{\jmath\left(\frac{-\beta^{2}+\alpha}{4 \alpha}\right)\pi}\Biggl[\operatorname{erf}\Bigl(\frac{e^{\jmath\frac{3\pi}{4}}\sqrt{\pi}(\beta-2\alpha)}{2\sqrt{\alpha}}\Bigr)
-\operatorname{erf}\Bigl(\frac{e^{\jmath\frac{3\pi}{4}}\sqrt{\pi}(\beta+2\alpha)}{2\sqrt{\alpha}}\Bigr)
\Biggr]}{4\sqrt{\alpha}}
.
\end{split}
\end{equation}
This completes the derivation.
\section{Derivation of Rice Distribution}\label{app:Rice}
The complex Gaussian pdf of \(y(\mathbf{v})\) is given by
\begin{equation}
f_{y(\mathbf{v})}(y) = \frac{1}{\pi\sigma^2}\exp\!\left(-\frac{|y - u e^{\jmath\psi}|^2}{\sigma^2}\right).
\end{equation}
Express \(y\) in polar form as \(y = z e^{\jmath \gamma}\), where \(z\ge0\) is the value of random variable $Z$ and \(\gamma\in[-\pi,\pi)\) is phase of $y$. With Jacobian \(z\), the joint pdf of \(z\) and \(\gamma\) is
\begin{equation}
\begin{split}
f_{z,\gamma}(z,\gamma) &= \frac{z}{\pi\sigma^2}\exp\!\left(-\frac{|z e^{\jmath \gamma} - u e^{\jmath\psi}|^2}{\sigma^2}\right)\\
     &= \frac{z}{\pi\sigma^2}\exp\!\left(-\frac{z^2+u^2-2zu\cos(\gamma-\psi)}{\sigma^2}\right).
     \label{eq:middle_rice}
\end{split}
\end{equation}
We can obtain the marginal pdf of \(z\) by integrating \Cref{eq:middle_rice} over \(\gamma\). By change of variable \(\phi=\gamma-\psi\), we have
\begin{equation}
f_z(z) = \frac{z}{\pi\sigma^2}\exp\!\left(-\frac{z^2+u^2}{\sigma^2}\right)
\int_{-\pi}^{\pi} \exp\!\left(\frac{2zu\cos\phi}{\sigma^2}\right)d\phi.
\end{equation}
Based on 
\begin{equation}
\int_{-\pi}^{\pi} e^{a\cos\phi}\,d\phi = 2\pi I_0(a),
\end{equation}
with \(a=\frac{2zu}{\sigma^2}\), we obtain the Rice distribution pdf
\begin{equation}
f_z(z) = \frac{2z}{\sigma^2}\exp\!\left(-\frac{z^2+u^2}{\sigma^2}\right)I_0\!\left(\frac{2zu}{\sigma^2}\right),\quad z\ge0.
\end{equation}
\printbibliography








\section*{Biography Section}
\vspace{-33pt}
\begin{IEEEbiographynophoto}{Zijun Wang} is currently a PhD student at Electrical Engineering department, State University of
New York at Buffalo, Buffalo NY USA. He received his B.E. degree in Telecommunications Engineering in 2024 at Nanjing University, Nanjing, China. His research interest is wireless communication.
\end{IEEEbiographynophoto}

\begin{IEEEbiographynophoto}
{Shawn Tsai} received the M.S.E.E. and Ph.D. degrees from Purdue University,
West Lafayette, IN, USA, in 1995 and 2000, respectively.
From 2000 to 2020, he was a System Engineer with Ericsson, Lewisville,
TX, USA; Huawei, Plano, TX, USA; and Qualcomm, San Diego, CA,
USA, contributing to research and development of wireless system standardization, design, implementation, and commercialization. He is currently
an Engineering Director of Communication System Design, MediaTek Inc.,
San Diego. His interests include communications theory, signal processing,
radio resource management, and air interface protocols.
\end{IEEEbiographynophoto}

\begin{IEEEbiographynophoto}
{Rama Kiran} received the M.Tech. degree in electrical engineering from the Indian Institute of Technology (IIT), Kanpur, in 2014, and the  Ph.D. degree in Electrical Communication Engineering from Indian Institute of Science (IISc), Bengaluru, in 2021. Currently, he is working as a research engineer at MediaTek (India) Pvt. Ltd., Bengaluru, where he is working on 5G/6G cellular standardization related research. From 2014 to 2015, he was with NI Systems (India) Pvt., Ltd., Bengaluru, where he worked on the development and implementation of algorithms for LTE and IEEE 802.11b/af/ah wireless standards. His research interests include wireless communication, reflective intelligent surfaces, and full-duplex communication.
\end{IEEEbiographynophoto}

\begin{IEEEbiographynophoto}{Rui Zhang}
 (Member, IEEE) is currently an Assistant professor at Electrical Engineering department, State University of
New York at Buffalo, Buffalo NY USA. She received a Ph.D.
degree at the School of Electrical and Computer Engineering,
Georgia Institute of Technology, Atlanta GA USA in 2022. In
2017, she received her B.S. degree in electrical engineering
and B.A. degree in economics both from Peking University,
Beijing, China. From 2022 to 2023, she was a staff engineer at
Communication System Design of MediaTek USA Inc. in San
Diego, CA. She is the recipient of the 2022 Marconi Young
scholar award and associate technical editor of IEEE Communication Magazine. Her research interests are wireless and
optical communications, and signal processing.
\end{IEEEbiographynophoto}

\vfill

\end{document}